\documentclass[12pt,preprint]{aastex}
\usepackage{mathrsfs, amsmath, amssymb, bm}

\usepackage{hyperref}

\begin{document}

\title{
    An N-body Integrator for Gravitating Planetary Rings,\\
    and the Outer Edge of Saturn's B Ring\vspace*{1.0in}
}

\author{
    Joseph M. Hahn
}
\affil{
    Space Science Institute\\
    c/o Center for Space Research\\
    University of Texas at Austin\\
    3925 West Braker Lane, Suite 200\\
    Austin, TX 78759-5378\\
    jhahn@spacescience.org\\
    512-992-9962\vspace*{1.0in}
}

\author{
    Joseph N.\ Spitale
}
\affil{
    Planetary Science Institute\\
    1700 East Fort Lowell, Suite 106\\
    Tucson, AZ 85719-2395\\
    jnspitale@psi.edu\\
    520-622-6300\vspace*{1.0in}
}

\author{
    Submitted for publication in the\\
    {\it Astrophysical Journal} on December 28, 2012\\
    Revised April 26, 2013\\
    Accepted June 1, 2013
    \ \vspace*{0.4in}
}

\begin{abstract}

A new symplectic N-body integrator is introduced, one designed to calculate
the global $360^\circ$ evolution of a self-gravitating planetary ring that is in orbit
about an oblate planet. This freely-available code is called {\tt epi\_int},
and it is distinct from other such codes in its use of streamlines to calculate
the effects of ring self-gravity. The great advantage of this approach is that
the perturbing forces arise from smooth wires of ring matter rather than discreet
particles, so there is very little gravitational scattering and so only a modest
number of particles are needed to simulate, say, the scalloped edge of a resonantly
confined ring or the propagation of spiral density waves.

The code is applied to the outer edge of Saturn's B ring, and a comparison
of Cassini measurements of the ring's forced response to simulations of Mimas' resonant
perturbations reveals that the B ring's surface density at its outer edge is 
$\sigma_0=195\pm 60$ gm/cm$^2$ which, if the same everywhere across the ring would
mean that the B ring's mass is about $90\%$ of Mimas' mass.

Cassini observations show that the B ring-edge has several free normal modes, which are
long-lived disturbances of the ring-edge that are not driven by any known satellite
resonances. Although the mechanism that excites or sustains these normal
modes is unknown, we can plant such a disturbance at a simulated ring's edge,
and find that these modes persist
without any damping for more than $\sim10^5$ orbits or $\sim100$ yrs
despite the simulated ring's viscosity $\nu_s=100$ cm$^2$/sec.
These simulations also indicate that impulsive disturbances at a ring can excite
long-lived normal modes, which suggests that an impact in the recent past
by perhaps a cloud of cometary debris might have excited these disturbances
which are quite common to many of Saturn's sharp-edged rings.

\end{abstract}

\section{Introduction}
\label{intro_section}

A planetary ring is often coupled dynamically to a satellite via orbital resonances.
The ring's response to resonant perturbations varies with the forcing, 
and if the ring is for instance composed of low optical depth dust, then the
ring's response will vary with the satellite's mass and its proximity. 
But in an optically thick planetary ring, 
such as Saturn's main A and B rings or its many dense narrow ringlets,
the ring is also interacting with itself via self gravity, 
so its response is also sensitive to the ring's
mass surface density $\sigma_0$ \citep{S84, MB05, HSP09}.
So by measuring a dense ring's response to satellite perturbations, and comparing 
that measurement to a model for the ring-satellite system, 
one can then infer the ring's physical properties, such as its surface
density $\sigma_0$, and perhaps other quantities too
\citep{MB05, TBN07, HSP09}. Recently \cite{HSP09}
developed a semi-analytic model of the outer edge of Saturn's B ring,
which is confined by an $m=2$ inner Lindblad resonance with the satellite Mimas.
The resonance index $m$ also describes the ring's anticipated equilibrium shape,
with the ring-edge's deviations from circular motion expected
to have an azimuthal wavenumber of $m=2$. So the B ring's expected 
shape is a planet-centered ellipse, which has $m=2$  alternating inward and
outward excursions. The model of \cite{HSP09} also calculates
the ring's equilibrium $m=2$ response excited by Mimas,
but that comparison between theory and observation was done during
the early days of the Cassini mission when that spacecraft's measurement
of the ring-edge's semimajor axis $a_{\mbox{\scriptsize edge}}$ was still
rather uncertain. It turns out that the ring's inferred surface density
is very sensitive to how far the B ring's outer edge extends beyond the resonance,
which was quite uncertain then due to the uncertainty in $a_{\mbox{\scriptsize edge}}$,
so the uncertainty in the ring's inferred $\sigma_0$ was also relatively large.
Now however $a_{\mbox{\scriptsize edge}}$ is known with much greater precision,
so a re-examination of this system is warranted.

\begin{figure}[th]
\epsscale{1.45}
\hspace*{-20ex}\plotone{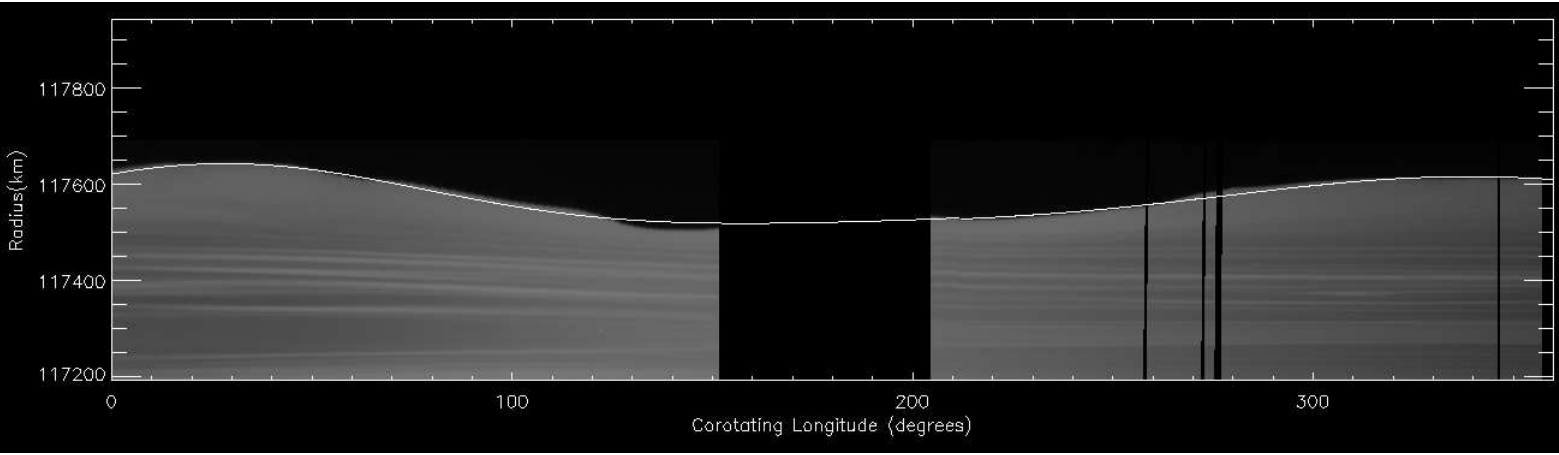}
\figcaption{
    \label{Bring_fig}
    A mosaic of Cassini images of the B ring's outer edge acquired during nine 
    hours on 28 January 208. Greyscale indicates the ring's optical surface
    brightness at various radii and corotating longitudes, meaning that
    local keplerian motion about an oblate
    planet is assumed as all the individual image elements
    are mapped to positions held at some common instant of 
    time. (Note thought that a true instantaneous snapshot of the ring would still
    have a different shape than this mosaic because the various normal modes rotate at
    differing speeds, and those differential rotations are
    not accounted for in this projection.)
    The curve at the ring's edge is the four-component kinematical model of \cite{SP10}, 
    which is a best fit to 18 such mosaics like this one but
    acquired over four years of monitoring, and the black zones are regions
    not used in that kinematic fit.
}
\end{figure}

\begin{figure}[th]
\epsscale{1.30}
\hspace*{-12ex}\plotone{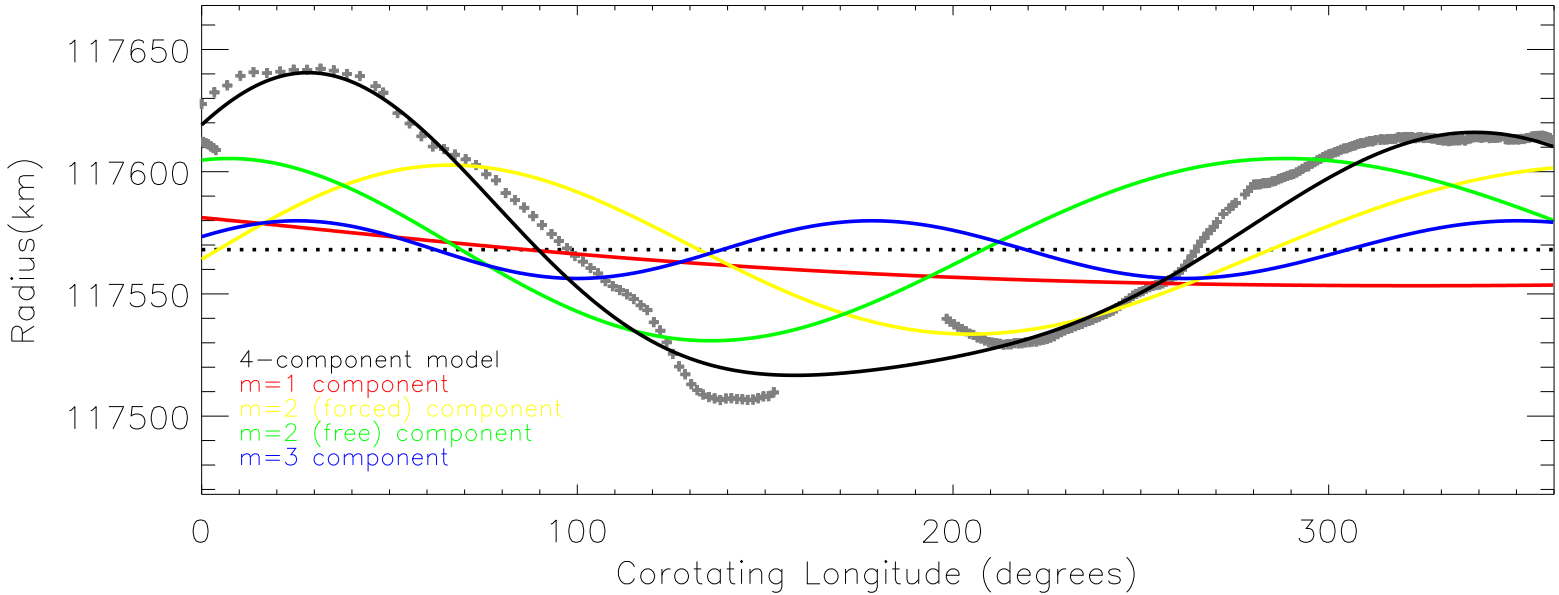}
\figcaption{
    \label{Bring_fit_fig}
    Crosses are the B ring-edge's observed radius versus corotating longitude on 
    28 January 2008
    extracted from the mosaic seen in Fig.\ \ref{Bring_fig}.
    Colored curves show the amplitudes and orientations the $m=1$, $m=2$ (forced), $m=2$ (free),
    and $m=3$ normal modes that \cite{SP10} fit to four years of Cassini imaging.
    Black curve is the superposition of those modes at this instant, and the dotted line is
    the B ring-edge's semimajor axis. Note
    that these curves do not agree at $0^\circ$ and $360^\circ$ corotating longitudes,
    due to the rotation of the normal modes that occurs during the nine hour
    observing window.
}
\end{figure}

Cassini's monitoring of the B ring also
reveals that the ring's outer edge exhibits several normal modes,
which are unforced disturbances that are not associated with any known
satellite resonances. Figure \ref{Bring_fig} illustrates this phenomenon with
a mosaic of images that Cassini acquired of the B ring's edge on 28 January 2008.
\cite{SP10} have also fit a kinematic model to four years worth of
Cassini images of the  B ring; that model is composed of four normal modes having azimuthal
wavenumbers $m=1,2,2,3$ that steadily rotate over time at distinct rates.
In the best-fitting kinematic model there are two $m=2$ modes,
one that is forced by and corotating with Mimas, as well as a free $m=2$ 
mode that rotates slightly faster. The amplitudes and orientations of all the modes
as they appear in the 28 January 2008 data is also shown in Fig.\ \ref{Bring_fit_fig}.
Note that although the B ring's outer edge, as seen in Fig.\ \ref{Bring_fig},
might actually resemble a simple $m=2$ shape on 28 January 2008,
at other times the ring-edge's shape is much more
complicated than a simple $m=2$ configuration, yet at other times the ring-edge is
relatively smooth and nearly circular; see for example Fig.\ 1 of \cite{SP10}.
This behavior is due to the superposition of the normal
modes that are rotating relative to each other,
which causes the B ring's edge to evolve over time.
Since this system is not in simple equilibrium, a time-dependent model of the
ring that does not assume equilibrium is appropriate here. 

So the following develops a new N-body method that is
designed specifically to track the time evolution of a self-gravitating
planetary ring, and that model is then applied to the latest Cassini results.   
Section \ref{method_section} describes in detail
the N-body model that can simulate all $360^\circ$ 
of a narrow annulus in a self-gravitating planetary ring
using a very modest number of particles.
Section \ref{B ring} then shows results from several simulations
of the outer edge of Saturn's B ring, and demonstrates how a ring's observed
epicyclic amplitudes and pattern speeds can be compared to N-body simulations
to determine the ring's physical properties. Results are then summarized
in Section \ref{summary}.


\section{Numerical method}
\label{method_section}

The following briefly summarizes the theory of the symplectic integrator
that \cite{DLL98} use in their {\tt SYMBA} code and \cite{C99} 
use in the {\tt MERCURY} integrator to calculate the motion 
of objects in nearly Keplerian orbits about a point-mass star. That numerical method
is adapted here so that one can study the evolution of
a self-gravitating planetary ring that is in orbit about an oblate planet.

\subsection{symplectic integrators}
\label{symplectic}

The Hamiltonian for a system of N bodies in orbit about a central planet is 
\begin{eqnarray}
    H &=& \sum_{i=0}^{N}\frac{p_i^2}{2m_i} + \sum_{i=0}^{N}\sum_{j>i}^{N} V_{ij},
\end{eqnarray}
where body $i$ has mass $m_i$ and momentum $\mathbf{p}_i = m_i\mathbf{v}_i$
where $\mathbf{v}_i=\mathbf{\dot{r}}_i$ is its velocity and $V_{ij}$
is the potential such that $\mathbf{f}_{ij}=-\nabla_{\mathbf{r}_i}V_{ij}$
is the force on $i$ due to body $j$ where $\nabla_{\mathbf{r}_i}$
is the gradient with respect to coordinate $\mathbf{r}_i$,
and the index $i=0$ is reserved for the central planet whose mass is $m_0$.
Next choose a coordinate system
such that all velocities are measured with respect to the system's barycenter,
so $\mathbf{p}_0 = -\sum_{j=1}^N\mathbf{p}_j$, and the Hamiltonian becomes
\begin{eqnarray}
    H &=& \sum_{i=1}^{N}\left(\frac{p_i^2}{2m_i} + V_{i0}\right) 
        + \sum_{i=1}^{N}\sum_{j>i}^NV_{ij}  
        + \frac{1}{2m_0}\left(\sum_{i=1}^N\mathbf{p}_i\right)^2 
        \equiv H_A + H_B + H_C
\end{eqnarray}
since $V_{ij} = V_{ji}$. This Hamiltonian has three parts,
\begin{mathletters}
    \label{H}
    \begin{eqnarray}
        H_A &=& \sum_{i=1}^N\left(\frac{p_i^2}{2m_i} + V_{i0}\right)\\
        H_B &=& \sum_{i=1}^N\sum_{j>i}^NV_{ij}\\
        H_C &=& \frac{1}{2m_0}\left(\sum_{i=1}^N\mathbf{p}_i\right)^2,
    \end{eqnarray}
\end{mathletters}
and the following will employ spatial coordinates such that all $\mathbf{r}_i$ are
measured relative to the central planet. 
This combination of planetocentric coordinates and barycentric
velocities is referred to as `democratic-heliocentric' coordinates in \cite{DLL98}
and `mixed-center' coordinates in \cite{C99}. 
In the above, $H_A$ is the sum of two-body Hamiltonians, 
$H_B$ represents the particles' mutual interactions, and
$H_C$ accounts for the additional forces that arise in this particular
coordinate system that are due to
the central planet's motion about the barycenter.

Hamilton's equations for the evolution of the coordinates $\mathbf{r}_i$
and momenta $\mathbf{p}_i$ for
particle $i\ge1$ are $\mathbf{\dot{r}}_i = \nabla_{\mathbf{p}_i}H$ and 
$\mathbf{\dot{p}}_i = -\nabla_{\mathbf{r}_i}H$. So
a particle that is subject only to Hamiltonian $H_B$ during short time interval
$\delta t$  would experience the velocity kick 
\begin{eqnarray}
    \label{dv}
    \mathbf{\delta v}_i &=& \frac{{\mathbf{\dot{p}}}_i \delta t}{m_i} = 
        -\nabla_{\mathbf{r}_i}H_B\frac{\delta t}{m_i} 
        = \frac{\delta t}{m_i}\sum_{j=1}^N\mathbf{f}_{ij},
\end{eqnarray}
which of course is $i$'s response to the forces exerted by
all the other small particles in the system.
And since $H_C$ is a function of momenta only, a particle subject to
$H_C$ during time $\delta t$ will see its spatial coordinate kicked by
\begin{eqnarray}
    \label{dx}
    \mathbf{\delta r}_i &=& \frac{\delta t}{m_0}\sum_{j=1}^{N} \mathbf{p}_j
\end{eqnarray}
due to the planet's motion about the barycenter.

Now let $\xi_i(t)$ represent any of particle $i$'s coordinates $x_i$ or
momenta $p_i$; that quantity evolves at the rate  \citep{G80}
\begin{eqnarray}
    \label{eom}
    \frac{d\xi_i}{dt} &=& [\xi_i, H] = [\xi_i, H_A + H_B + H_C]
        =(A + B + C) \xi_i
\end{eqnarray}
where the brackets are a Poisson bracket,
and the operator $A$ is defined such that $A\xi_i = [\xi_i, H_A]$, with
operators $B$ and $C$ defined similarly. The solution to Eqn.\ (\ref{eom})
for $\xi_i$ evaluated at the later time $t + \Delta t$ is formally
\begin{eqnarray}
    \label{eom_soln}
    \xi_i(t + \Delta t) &=& e^{(A + B + C)\Delta t}\xi_i(t)
\end{eqnarray}
\citep{G80}, but this exact expression is in general not analytic and not in a useful
form. However \cite{DLL98} and \cite{C99} show that the above is approximately
\begin{eqnarray}
    \label{eom_soln_approx}
    \xi_i(t + \Delta t) &\simeq& e^{B\Delta t/2}e^{C\Delta t/2}e^{A\Delta t}
        e^{C\Delta t/2}e^{B\Delta t/2}\xi_i(t),
\end{eqnarray}
which indicates that five actions that are to occur as 
the system of orbiting bodies are advanced one timestep $\Delta t$ by the integrator.
First ({\it i.}) the operator $e^{B\Delta t/2}$ acts on
$\xi_i(t)$, which increments ({\it i.e.} kicks) particle $i$'s
velocity $\mathbf{v}_i$ by Eqn.\ (\ref{dv}) due to the system's interparticle forces
with $\delta t=\Delta t/2$. Then ({\it ii.})
the  $e^{C\Delta t/2}$ operator acts on the result of substep ({\it i.})
and kicks the particle's spatial coordinates $\mathbf{r}_i$ according to
Eqn.\ (\ref{dx}) due to the central planet's motion about the barycenter. Then
in substep ({\it iii.}) the $e^{A\Delta t}$ operation advances
the particle along its unperturbed epicyclic orbit about the central planet during a full
timestep $\Delta t$, with
this substep is referred to below as the orbital `drift' step. Step ({\it iv.})
is another coordinate kick $\delta\mathbf{r}_i$ and the last step ({\it v.})
is the final velocity kick. 

In a traditional symplectic N-body integrator the planet's oblateness
is treated as a perturbation whose effect would be accounted for
during steps ({\it i.})\ and ({\it v.})\ which provide an extra
kick to a particle's velocity every timestep. Those kicks
cause a particle in a circular orbit to have a tangential speed
that is faster than the Keplerian speed by the fractional amount
that is of order $\sim J_2(R/r)^2\sim3\times10^{-3}$ where $J_2\simeq0.016$ is
Saturn's second zonal harmonic and $r/R\sim2$ is a B ring particle's orbit radius $r$
in units of Saturn's radius $R$. The particle's circular speed is super-Keplerian,
and if its coordinates and velocities were to be converted to Keplerian orbit elements,
its Keplerian eccentricity would also be of order $e\sim3\times10^{-3}$.
This putative eccentricity should be compared
to the observed eccentricity of Saturn's B ring, which is the focus of this
study and is of order $e\sim10^{-4}$,
about 30 times smaller than the particle's Keplerian eccentricity.
The main point is, that one does not want to use Keplerian orbit elements
when describing a particle's nearly circular motions about an oblate planet
because the Keplerian eccentricity is dominated by planetary oblateness whose effects
obscures the ring's much smaller forced motions. 

To sidestep this problem, the following algorithm uses the {\em epicyclic} orbit
elements of \cite{BL94} which provide a more accurate representation
of an unperturbed particle's orbit about 
an oblate planet. Note that this use of epicyclic orbit elements
effectively takes the effects of oblateness out of the 
integrator's velocity kick steps ({\it i.})\ and ({\it v.}) and places
oblateness effects in the integrator's drift step ({\it iii.}), which is preferable
because the forces in the B ring that are due to oblateness are about
$\sim10^4$ times larger than any satellite perturbation. 
The following details how these epicyclic orbit elements are calculated
and are used to evolve the particle along its unperturbed orbit during the drift
substep.

\subsection{epicyclic drift}
\label{drift}

This 2D model will track a particle's motions in the ring plane, so 
the particle's position and velocity relative to the central planet
can be described by four epicyclic orbit elements:
semimajor axis $a$, eccentricity $e$,
longitude of periapse $\tilde{\omega}$, and mean anomaly $M$. For a particle
in a low eccentricity orbit about an oblate planet, the 
relationship between the particle's epicyclic orbit elements
and its cylindrical coordinates $r, \theta$ and velocities $v_r, v_\theta$ are
\begin{mathletters}
    \label{rv}
    \begin{eqnarray}
        \label{r}
        r &=& a\left[1 - e\cos M +
            \left(\frac{\eta_0}{\kappa_0}\right)^2(2 - \cos^2M)e^2 \right] \\
        \theta &=& \tilde{\omega} + M + 
            \frac{\Omega_0}{\kappa_0}\left\{2e\sin M + \left[\frac{3}{2} + 
            \left(\frac{\eta_0}{\kappa_0}\right)^2\right]e^2\sin M\cos M\right\}\\
        v_r &=& a\kappa_0\left[e\sin M + 
            2\left(\frac{\eta_0}{\kappa_0}\right)^2e^2\sin M\cos M\right]\\
        \label{v_t}
        v_\theta &=& a\Omega_0\left\{1 + e\cos M - 
            2\left(\frac{\eta_0}{\kappa_0}\right)^2e^2 + 
            \left[1 + \left(\frac{\eta_0}{\kappa_0}\right)^2\right]e^2\cos^2 M\right\},
    \end{eqnarray}
\end{mathletters}
which are adapted from Eqns.\ (47-55) of \cite{BL94}. These equations are accurate to
order ${\cal O}(e^2)$ and require $e\ll1$.
Here $\Omega_0(a)$ is the angular velocity of a particle in a circular
orbit while $\kappa_0(a)$
is its epicyclic frequency and the frequency $\eta_0(a)$
is defined below, all of which are functions of the particle's semimajor axis $a$.
Also keep in mind that when the following
refers to the particle's orbit elements, it is the
{\em epicyclic} orbit elements that are intended\footnote{Actually what we identify here
as the semimajor axis $a$ is called $r_0$ in \cite{BL94},
which differs slightly from what they identify as the epicyclic semimajor axis
$a_e$ where $a_e=r_0(1 + e^2)$.}, which are distinct
from the {\em osculating} orbit elements that describe
pure Keplerian motion around a spherical planet.
But these distinctions disappear in the limit that the planet becomes
spherical and the orbit frequencies $\Omega_0, \kappa_0$, and $\eta_0$
all converge on the mean motion $\sqrt{Gm_0/a^3}$, where $G$ is 
the gravitational constant and $m_0$ is the central planet's mass;
in that case, Eqns.\ (\ref{rv}) recover a Keplerian orbit to order ${\cal O}(e^2)$.

The three orbit frequencies $\Omega_0$, $\kappa_0$, and $\eta_0$
appearing in Eqns.\ (\ref{rv})
are obtained from spatial derivatives of the oblate planet's
gravitational potential $\Phi$, which is
\begin{eqnarray}
    \Phi(r) &=& -\frac{Gm_0}{r} + 
        \frac{Gm_0}{r}\sum_{k=1}^{\infty}J_{2k}P_{2k}(0)
        \left(\frac{R_p}{r}\right)^{2k}
\end{eqnarray}
where $R_p$ is the planet's effective radius, $J_{2k}$ is one of the oblate
planet's zonal harmonics, and $P_{2k}(0)$ is a Legendre polynomial with zero argument. 
For reasons that will be evident shortly, these calculations
will only preserve the $J_{2}$ term in the above sum, so
\begin{eqnarray}
    \Phi(r) &=& -\frac{Gm_0}{r}\left[ 1 + 
        \frac{1}{2}J_2\left(\frac{R_p}{r}\right)^2\right]
\end{eqnarray}
and the orbital frequencies are
\begin{mathletters}
    \label{orbit_frequencies}
    \begin{eqnarray}
        \label{Omega^2}
        \Omega_0^2(a) &=& \left.\frac{1}{r}\frac{\partial\Phi}{\partial r}\right|_{r=a}
            = \frac{Gm_0}{a^3}\left[1 + \frac{3}{2}J_2\left(\frac{R_p}{a}\right)^2 \right]\\
        \label{kappa^2}
         \kappa_0^2(a) &=& \left.\frac{3}{r}\frac{\partial\Phi}{\partial r}\right|_{r=a}
            + \left.\frac{\partial^2\Phi}{\partial r^2}\right|_{r=a}
            = \frac{Gm_0}{a^3}\left[1 - \frac{3}{2}J_2\left(\frac{R_p}{a}\right)^2 \right]\\
        \eta_0^2(a) &=& \left.\frac{2}{r}\frac{\partial\Phi}{\partial r}\right|_{r=a}
            - \left.\frac{r}{6}\frac{\partial^3\Phi}{\partial r^3}\right|_{r=a}
            = \frac{Gm_0}{a^3}\left[1 - 2J_2\left(\frac{R_p}{a}\right)^2 \right]\\
        \beta_0^2(a) &=& - \left.\frac{r^4}{24}\frac{\partial^4\Phi}{\partial r^4}\right|_{r=a}
            = \frac{Gm_0}{a^3}\left[1 + \frac{15}{2}J_2\left(\frac{R_p}{a}\right)^2 \right]
    \end{eqnarray}
\end{mathletters}
where the additional frequency $\beta_0(a)$ is needed below.

During the particle's unperturbed epicyclic drift phase its angular orbit elements
$M$ and $\tilde{\omega}$ advance during timestep $\Delta t$ by amount
\begin{mathletters}
    \label{dM}
    \begin{eqnarray}
        \Delta M &=& \kappa\Delta t\\
        \Delta \tilde{\omega} &=& (\Omega - \kappa)\Delta t
    \end{eqnarray}
\end{mathletters} 
where the frequencies $\Omega$ and $\kappa$ in Eqns.\ (\ref{dM}) differ slightly
from Eqns.\ (\ref{orbit_frequencies})  due to additional corrections
that are of order ${\cal O}(e^2)$:
\begin{mathletters}
    \label{Omega_kappa}
    \begin{eqnarray}
        \Omega(a,e) &=& \Omega_0\left\{1 + 
           3\left[\frac{1}{2} - \left(\frac{\eta_0}{\kappa_0}\right)^2\right]e^2 \right\}\\
        \kappa(a,e) &=& \kappa_0\left(1 + 
            \left\{\frac{15}{4}\left[\left(\frac{\Omega_0}{\kappa_0}\right)^2 
            - \left(\frac{\eta_0}{\kappa_0}\right)^4\right] - 
            \frac{3}{2}\left(\frac{\beta_0}{\kappa_0}\right)^2\right\}e^2 \right)
    \end{eqnarray}
\end{mathletters} 
\citep{BL94}.

\cite{BL94} also show that the above equations have three integrals
of the motion: the particle's specific energy $E$, its specific angular momentum $h$, and 
its epicyclic energy $I_3$. Those integrals are
\begin{mathletters}
    \label{integrals}
    \begin{eqnarray}
        E &=& \frac{1}{2}(v_r^2 + v_\theta^2) + \Phi(r) = \frac{1}{2}(a\Omega_0)^2 + \Phi(a)
        + \frac{1}{2}(a\kappa_0)^2 e^2 + {\cal O}(e^4)\\
        h &=& r v_\theta = a^2\Omega_0 + {\cal O}(e^4)\\
        \label{I3}
        \mbox{and}\quad I_3 &=& \frac{1}{2}[v_r^2 + \kappa_0^2(r-a)^2] - \eta_0^2(r-a)^3/a
            = \frac{1}{2}(a\kappa_0e)^2  + {\cal O}(e^4).
    \end{eqnarray}
\end{mathletters} 

Advancing the particle along its epicyclic orbit require converting
its cylindrical coordinates and velocities into epicyclic orbit elements.
To obtain the particle's semimajor axis, solve the angular momentum
integral $h(a)=a^2\Omega_0$, which is quadratic in $a$ so
\begin{eqnarray}
    \label{a}
    a &=&g\left(1 + \sqrt{1 - \frac{3J_2}{2g^2}}\right) R_p
\end{eqnarray}
where $g=(rv_\theta)^2/2Gm_0R_p$. Note though that if the $J_4$
and higher oblateness terms had been preserved
in the planet's potential, then the angular momentum
polynomial would be of degree 4 and higher in $a$, for which there is no known
analytic solution. That equation could still be solved numerically,
but that step would have to be performed for all particles at every timestep,
which would slow the N-body algorithm so much as to make it useless.
So only the $J_2$ term is preserved here, which
nonetheless accounts for the effects of
planetary oblateness in a way that is sufficiently realistic.

To calculate the particle's remaining orbit elements, use Eqn.\ (\ref{I3}) to obtain
the $I_3$ integral which then provides its eccentricity via
\begin{eqnarray}
    \label{e}
    e &=& \frac{\sqrt{2I_3}}{a\kappa_0}.
\end{eqnarray}
Then set $x=e\cos M$ and $y=e\cos M$ and solve Eqns.\ (\ref{r}) and (\ref{v_t}) 
for $x$ and $y$:
\begin{mathletters}
    \label{xy}
    \begin{eqnarray}
        x &=& \left(\frac{\eta_0}{\kappa_0}\right)^2
            \left[2(1+e^2) - \frac{v_\theta}{a\Omega_0} - \frac{r}{a}\right] 
            + 1 - \frac{r}{a}\\
        \mbox{and}\quad y &=& \frac{v_r/a\kappa_0}{1 + 2(\eta_0/\kappa_0)^2x},
    \end{eqnarray}
\end{mathletters} 
which then provides the mean anomaly via $\tan M=y/x$.

To summarize, the epicyclic drift step uses Eqns.\ (\ref{integrals}--\ref{xy})
to convert each particle's cylindrical coordinates
into epicyclic orbit elements. The particles' orbit frequencies 
$\Omega(a,e)$ and $\kappa(a,e)$ are obtained via Eqns.\ (\ref{orbit_frequencies}) and
(\ref{Omega_kappa}), and Eqns.\ (\ref{dM}) are then used to
advance each particle's orbit elements $M$ and $\tilde{\omega}$
during timestep $\Delta t$, with Eqns.\ (\ref{rv})
used to convert the particles' orbit elements back into cylindrical coordinates.

\subsection{velocity kicks due to the ring's internal forces}
\label{kicks}

The N-body code developed here is designed to follow the dynamical
evolution of all $360^\circ$ of a narrow annulus within a planetary
ring, and it is intended to deliver accurate results quickly using a desktop PC.
The most time consuming part of this algorithm is the calculation
of the accelerations that the gravitating ring exerts on all of its particles, 
so the principal goal here is to design an algorithm that will calculate
these accelerations with sufficient accuracy while using the fewest possible
number of simulated particles.


\subsubsection{streamlines}
\label{streamlines}

The dominant internal force in a dense planetary ring is its self gravity,
and the representation of the ring's full $360^\circ$ extent
via a modest number of {\em streamlines}
provides a practical way to calculate rapidly the acceleration that the entire ring
exerts on any one particle. A streamline is the closed path through the ring
that is traced by those particles that share a common initial semimajor axis $a$.
The simulated portion of the planetary ring will be
comprised of $N_r$ discreet streamlines that are spaced evenly in semimajor axis $a$,
with each streamline comprised of $N_\theta$ particles
on each streamline, so a model ring consists of $N_rN_\theta$
particles. Simulations typically employ $N_r\sim100$ streamlines 
with $N_\theta\sim50$ particles along each streamline, 
so a typical ring simulation uses about five thousand particles. 
Note though that the assignment of particles to a given streamline
is merely labeling;
particles are still free to wander over time in response to the
ring's internal forces: gravity, pressure, and viscosity.
But as the following will show, the simulated ring stays coherent and highly organized
throughout the run, in the sense that particles
on the same streamline do not pass each other longitudinally,
nor do adjacent streamlines cross. Because the simulated ring
stays so highly organized, there is no radial
or transverse mixing of the ring particles, and
the particles will preserve over time membership in their
streamline\footnote{But if the simulated ring is instead initialized
with all particles on a given streamline having distinct (rather than
common) values for $a$ and $e$, then the resulting streamlines 
can appear ragged in
longitude $\theta$. And if that initial ring is sufficiently
ragged or non-smooth, then that raggedness can
grow over time as the particles $a$'s and $e$'s evolve independently. The main
point is that the streamline model employed here succeeds when
all streamlines are sufficiently smooth, and that is accomplished
by initializing all particles in a given streamline with commmon $a,e$.}.

\subsubsection{ring self gravity}
\label{ring_gravity}

\begin{figure}[th]
\epsscale{1.0}
\vspace*{-4ex}\plotone{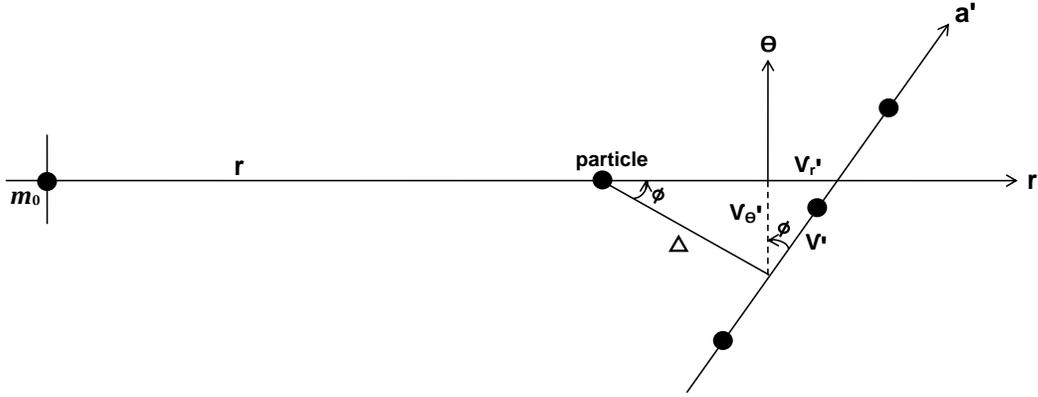}\vspace*{-14ex}
\figcaption{
    \label{streamline_fig}
    A particle lies a distance $r$ from the central mass $m_0$ and is
    perturbed by a streamline whose particles have semimajor axes $a'$. The shape
    of that streamline is determined by fitting a Lagrange
    polynomial to the three particles that are nearest in longitude,
    which is represented by the nearly straight curve $a'$,
    with that polynomial then providing the streamline's distance $\Delta$
    from the particle at $r$. The streamline's gravitational acceleration of
    that particle is $A_g=2G\lambda/\Delta$, which has radial and tangential
    components $A_{g,r}=A_g\cos\phi$ and $A_{g,\theta}=-A_g\sin\phi$
    where angle $\phi$ obeys $\sin\phi=v_r'/v'\simeq v_r'/v_\theta'$ and
    $\cos\phi=v_\theta'/v'\simeq1$ to lowest order in the perturbing streamline's
    eccentricity $e'$, so $A_{g,r}\simeq A_g$ and 
    $A_{g,\theta}\simeq-A_gv_r'/v_\theta'$.\vspace*{2ex}
}
\end{figure}

The concept of gravitating streamlines is widely used in analytic studies
of ring dynamics \citep{GT79, BGT83aj, BGT86, LR95, HSP09}, and the concept is easily
implemented in an N-body code. 
Because the simulated portion of the ring is narrow, 
its streamlines are all close in the
radial sense. Consequently the
gravitational pull that one streamline exerts on a particle is dominated by the nearest
part of the streamline, with that acceleration being quite insensitive
to the fact that the more distant and unimportant parts of the perturbing
streamline are curved. So the perturbing streamline can be regarded as
a straight and infinitely long wire of matter whose linear density is
$\lambda\simeq m_pN_\theta /2\pi a$ to lowest order in the streamline's small eccentricity
$e$, where $m_p$ is the mass of a single particle. The gravitational acceleration that a
wire of matter exerts on the particle is 
\begin{eqnarray}
    \label{Ag}
    A_g &=& \frac{2G\lambda}{\Delta}
\end{eqnarray}
where $\Delta$ is the separation between the particle and the streamline.
However the particles in that streamline only provide $N_\theta$ discreet samplings
of a streamline that is after all slightly curved over larger spatial scales.
So to find the distance to nearest part of the perturbing streamline,
the code identifies at every timestep the three perturbing particles that are nearest
in longitude to the perturbed particle. A second-degree Lagrange polynomial is then used
to fit a smooth continuous curve through those three particles \citep{KS}, and this
polynomial provides a convenient method for extrapolating the 
perturbing streamline's distance $\Delta$ from the perturbed particle. This procedure
is also illustrated in Fig.\ \ref{streamline_fig}, which shows that the
radial and tangential components of that acceleration are
\begin{mathletters}
    \label{Ag_r_theta}
    \begin{eqnarray}
        A_{g,r} &\simeq& A_g\\
        \mbox{and}\qquad A_{g,\theta} &\simeq& -A_g v_r'/v_\theta'
    \end{eqnarray}
\end{mathletters} 
to lowest order in the perturbing streamline's eccentricity $e'$,
where $v_r'$ and $v_\theta'$ are the radial and tangential velocity components of
that streamline. Equation (\ref{Ag_r_theta}) is then summed to obtain 
the gravitational acceleration that all other streamlines exerts on the particle.

To obtain the gravity that is exerted by the streamline that the particle inhabits,
treat the particle as if it resides in a gap in that streamline that extends midway
to the adjacent particles. The nearby portions of that streamline can be regarded
as two straight and semi-infinite lines of matter pointed at the particle
whose net gravitational acceleration is
\begin{eqnarray}
    \label{Ag_streamline}
    A_g &=& 2G\lambda\left(\frac{1}{\Delta_+} - \frac{1}{\Delta_-}\right)
\end{eqnarray}
where $\Delta_+$ and $\Delta_-$ are the particle's distance from its neighbors
in the leading (+) and trailing (-) directions. The radial and tangential
components of that streamline's gravity are
\begin{mathletters}
    \label{Ag_r_theta_streamline}
    \begin{eqnarray}
        A_{g,r} &\simeq& A_g v_r/v_\theta\\
        \mbox{and}\qquad A_{g,\theta} &\simeq& A_g
    \end{eqnarray}
\end{mathletters} 
where $v_r, v_\theta$ are the perturbed particle's velocity components.

A major benefit of using Eqn.\ (\ref{Ag}) to calculate the ring's gravitational
acceleration is that there is no artificial gravitational stirring. This is in
contrast to a traditional N-body model that would use discreet point masses
to represent what is really a continuous ribbon of 
densely-packed ring matter. Those gravitating point masses then tug
on each other in amounts that very rapidly in magnitude and direction
as they drift past each other in longitude, and
those rapidly varying tugs will quickly excite the simulated particles'
dispersion velocity. As a result, the particles' unphysical
random motions tend to wash out the ring's large-scale coherent forced motions,
which is usually the quantity that is of interest.
So, although Eqn.\ (\ref{Ag}) is only approximate because it does not
account for the streamline's curvature that occurs far away from a perturbed ring
particle, Eqn.\ (\ref{Ag}) is still much more realistic and accurate
than the force law that would be employed in a traditional global 
N-body simulation of a planetary ring, which
out of computational necessity would treat a continuous stream of
ring matter as discreet clumps of overly massive
gravitating particles.

\subsubsection{ring pressure}
\label{pressure}

A planetary ring is very flat and its vertical structure will be unresolved in this
model, so a 1D pressure $p$ is employed here. That pressure $p$ is the
rate-per-length that a streamline segment communicates linear momentum
to the adjacent streamline orbiting just exterior to it, with that momentum
exchange being due to collisions occurring among particles
on adjacent streamlines. So for a small streamline
segment of length $\delta\ell$ that resides somewhere in the ring's interior, the
net force on that segment due to ring pressure is 
$\delta f = [p(r-\Delta) - p(r)]\delta\ell$ since $p(r-\Delta)$ is the
pressure or force-per-length exerted by the streamline that lies just
interior and a distance $\Delta$ away from segment $\delta\ell$, 
and $p(r)$ is the force-per-length
that segment $\delta\ell$ exerts on the exterior streamline. And since
force $\delta f = A_p \delta m$ where 
$\delta m=\lambda\delta\ell$ is the segment's mass,
the acceleration on a particle due to ring pressure is
\begin{eqnarray}
    \label{Ap}
    A_p &=& \frac{\delta f}{\delta m} = \frac{p(r-\Delta) - p(r)}{\lambda}
        \simeq-\frac{\Delta}{\lambda}\frac{\partial p}{\partial r}
        =-\frac{1}{\sigma}\frac{\partial p}{\partial r}
\end{eqnarray}
since the ring's surface density $\sigma=\lambda/\Delta$.

Formulating the acceleration in terms of pressure differences across adjacent
streamlines is handy because the 
model can then easily account for the large pressure drop that
occurs at a planetary ring's edge, which can be quite
abrupt when the ring's edge is sharp.
For a particle orbiting at the ring's innermost streamline, 
the acceleration there is simply $A_p=-p(r)/\lambda$
since there is no ring matter orbiting interior to it so $p(r-\Delta)=0$ there; likewise
the acceleration of a particle in the ring's outermost streamline is
$A_p = p(r-\Delta)/\lambda$. Pressure is exerted perpendicular
to the streamline and hence it is predominantly a radial force,
so by the geometry of Fig.\ \ref{streamline_fig} the radial
component of the acceleration due to pressure is $A_{p,r}\simeq A_p$
while the tangential component $A_{p,\theta}\simeq-A_pv_r/v_\theta$
is smaller by a factor of $e$,
where $v_r$ and $v_\theta$ are the perturbed particle's radial and tangential
velocities. This accounts for the pressure on the particle due to
adjacent streamlines.

The acceleration on the particle due to pressure gradients in the
particle's streamline is simply $A_p=-(\partial p/\partial\theta)/(r\sigma)$. This
acceleration points in the direction of the particle's motion,
so the radial and tangential components of that acceleration are
$A_{p,r}\simeq A_p v_r/v_\theta$ and $A_{p,\theta}\simeq A_p$.

Acceleration due to pressure requires selecting an equation of state (EOS)
that relates the pressure $p$ to the ring's other properties, and this study
will treat the ring as a dilute gas of colliding particles
for which the 1D pressure is $p=c^2\sigma$ where $c$ is the
particles dispersion velocity. However alternate EOS exist for planetary
rings, and that possibility is discussed in Section \ref{EOS}.

A simple finite difference scheme is used to calculate the pressure gradient in
Eqn.\ (\ref{Ap}) in the vicinity of particle $i$ in streamline $j$ that lies at
at longitude $\theta_{i,j}$. Lagrange polynomials are again used 
to evaluate the adjacent streamlines' planetocentric distances $r_{i, j-1}$ and
$r_{i, j+1}$ along the particle's longitude $\theta_{i,j}$, 
so the pressure gradient at particle $i$ in streamline $j$ is
\begin{eqnarray}
    \label{dp_dr}
    \left.\frac{\partial p}{\partial r}\right|_{i,j} &\simeq& 
        \frac{p_{i,j+1} - p_{i, j-1}}{r_{i, j+1} - r_{i, j-1}}
\end{eqnarray}
where the pressures in the adjacent streamlines $p_{i, j+1}$ and $p_{i, j-1}$
are also determined by interpolating those quantities to 
the perturbed particle's longitude $\theta_{i,j}$.

The surface density $\sigma_{i,j}$
in the vicinity of particle $i$ in streamline $j$ is determined by centering a box
about that particle whose radial extent spans half the distance to
the neighboring streamlines, so
\begin{eqnarray}
    \label{sigma}
    \sigma_{i,j} &=& \frac{2\lambda_j}{r_{i,j+1} - r_{i,j-1}}.
\end{eqnarray}
If however streamline $j$ lies at the ring's inner edge where $j=0$ 
then the surface density
there is $\sigma_{i,0}=\lambda_0/(r_{i,1} - r_{i,0})$
while the surface density at the outermost $j=N_r -1$ streamline is
$\sigma_{i, N_r - 1}=\lambda_{N_r - 1}/(r_{i, N_r - 1} - r_{i, N_r - 2})$.

\subsubsection{ring viscosity}
\label{viscosity}

Viscosity has two types, shear viscosity and bulk viscosity.
Shear viscosity is the friction
that results as particles on adjacent streamlines collide as they flow past each 
other.  The friction due to this shearing motion
causes adjacent streamlines to torque each other, so shear
viscosity communicates a radial flux of angular momentum through the ring.
A particle on a streamline experiences a net torque and hence a tangential acceleration
when there is a radial gradient in that angular momentum flux.

And if there are additional
spatial gradients in the ring's velocities that cause ring particles to converge
towards or diverge away from each other, then these relative motions will
cause ring particles to bump each other as they flow past, which transmits momentum
through the ring via the pressure forces discussed above. However
the ring particles' viscous bulk friction tends to retard
those relative motions, and that friction results in an additional flux of
linear momentum through the ring. Any radial gradients in that linear momentum
flux then results in a radial acceleration on a ring particle.

The 1D radial flux of the $z$ component of angular momentum due to the ring's
shear viscosity is derived in Appendix \ref{shear_appendix}:
\begin{eqnarray}
    \label{F_shear}
    F &=& -\nu_s\sigma r^2\frac{\partial\dot{\theta}}{\partial r}
\end{eqnarray}
(see Eqn.\ \ref{F_app})
where $\nu_s$ is the ring's kinematic shear viscosity and $\dot{\theta}=v_\theta/r$
is the angular velocity.
The quantity $F$ is the rate-per-length that one streamline
segment communicates angular momentum to the adjacent streamline orbiting just exterior,
so the net torque on a streamline segment of length $\delta\ell$ is
$\delta\tau=[F(r-\Delta) - F(r)]\delta\ell$ but 
$\delta\tau=rA_{\nu,\theta}\delta m$ where $\delta m = \lambda\delta\ell$
so the tangential acceleration due to the ring's shear viscosity is
\begin{eqnarray}
    \label{A_vs}
    A_{\nu,\theta} &=& \frac{F(r-\Delta) - F(r)}{\lambda r}
        =-\frac{1}{\sigma r}\frac{\partial F}{\partial r}.
\end{eqnarray}
Again this differencing approach is useful because it easily accounts
for the large viscous torque that occurs at a ring's sharp edge since
$A_{\nu,\theta}=- F(r)/\lambda r$ at the ring's inner edge and
$A_{\nu,\theta}=F(r - \Delta)/\lambda r$ at the ring's outer edge.

Appendix \ref{bulk_appendix} shows that the radial flux of linear momentum due to the
ring's shear and bulk viscosity is
\begin{eqnarray}
    \label{G}
    G &=& -\left(\frac{4}{3}\nu_s + \nu_b\right)\sigma\frac{\partial v_r}{\partial r}
        - \left(\nu_b - \frac{2}{3}\nu_s\right)\frac{\sigma v_r}{r}
\end{eqnarray}
(Eqn.\ \ref{G_appendix2})
where $\nu_b$ is the ring's bulk viscosity. This quantity is analogous to
a 1D pressure so the corresponding acceleration is
\begin{eqnarray}
    \label{A_vb}
    A_{\nu,r} &=& \frac{G(r-\Delta) - G(r)}{\lambda}
        =-\frac{1}{\sigma}\frac{\partial G}{\partial r}
\end{eqnarray}
in the ring's interior and $A_{\nu,r}=-G(r)/\lambda$ or
$A_{\nu,r}=G(r-\Delta)/\lambda$ along the ring's inner or outer edges.

To evaluate the partial derivatives that appear in the flux equations (\ref{F_shear})
and (\ref{G}), Lagrange polynomials are again used to determine the
angular and radial velocities $\dot{\theta}$ and $v_r$ in the adjacent
streamlines, interpolated at the perturbed particle's longitude, with
finite differences used to calculate the radial gradients in those quantities.

\subsubsection{satellite gravity}
\label{sat_gravity}

All ring particles are also subject to each satellite's gravitational
acceleration, $A_s=Gm_s/\Delta^2$, where $m_s$ is the satellite's mass
and $\Delta$ is the particle-satellite separation. Satellites also feel
the gravity exerted by all the ring particles, as well as the satellites' mutual
gravitational attractions.

And once all of the accelerations of each ring particle and satellite are tallied,
each body is then subject to the corresponding velocity kicks of
steps ({\it i.}) and ({\it v.}) 
that are described just below Eqn.\ (\ref{eom_soln_approx}).

\subsection{tests of the code}
\label{tests}

The N-body integrator developed here is called {\tt epi\_int},
which is shorthand for {\em epicyclic integrator}, and the following briefly
describes the suite of simulations whose known outcomes are used to
test all of the code's key parts.

{\bf Forced motion at a Lindblad resonance:} numerous
massless particles are placed in circular orbits at Mimas' $m=2$
inner Lindblad resonance. In this test, Mimas' initially zero mass is 
slowly grown to its current mass over an exponential
timescale $\tau_s=1.6\times10^4$ ring orbits,
which excites adiabatically the ring particle's forced eccentricities to levels
that are in excellent agreement with the solution to the linearized equations
of motion, Eqn.\ (42) of \cite{GT82}. Similar results are also obtained
for the particle's response to Janus' $m=7$ inner Lindblad resonance,
which is responsible for confining the outer edge of Saturn's A ring. These simulations test
the implementation of the integrator's kick-step-drift scheme as well as the satellite's
forcing of the ring.

{\bf Precession due to oblateness:} this simple test confirms that the orbits of
massless particles in low eccentricity orbits precess at the expected
rate, $\dot{\tilde{\omega}}(a)=\Omega - \kappa = \frac{3}{2}J_2(R_p/a)^2\Omega(a)$, 
due to planetary oblateness $J_2$.

{\bf Ringlet eccentricity gradient and libration:} when a narrow eccentric ringlet
is in orbit about an oblate planet, dynamical equilibrium requires the ringlet
to have a certain eccentricity gradient so that differential
precession due to self-gravity cancels that due to oblateness. And
when the ringlet is composed of only two streamlines then
this scenario is analytic, with the ringlet's equilibrium eccentricity
gradient given by Eqn.\ (28b) of \cite{BGT83}. So to test {\tt epi\_int}'s treatment
of ring self-gravity, we perform a suite of simulations of narrow eccentric
ringlets that have surface densities $40<\sigma<1000$ gm/cm$^2$ with initial
eccentricity gradients given by Eqn.\ (28b), and integrate over time
to show that these pairs of streamlines do indeed precess in sync with no relative
precession, as expected, over runtimes that exceed of the timescale
for massless streamlines to precess differentially.
And when we repeat these experiments with the ringlets
displaced slightly from their equilibrium eccentricity gradients, we find
that the simulated streamlines librate at the frequency given by Eqn.\ (30)
of \cite{BGT83}, as expected.

{\bf Density waves in a pressure-supported disk:} this test examines
the model's treatment of disk pressure, and uses a satellite to launch a two-armed
spiral density wave at its $m=2$ ILR in a non-gravitating pressure supported
disk. The resulting pressure wave has a wavelength
and amplitude that agrees with Eqn.\ (46) of \cite{W86}, as expected. 

{\bf Viscous spreading of a narrow ring:} in this test
{\tt epi\_int} follows the radial evolution of an initially narrow
ring as it spreads radially due to its viscosity, and
the simulated ring's surface density $\sigma(r,t)$ is in excellent agreement with the
expected solution, Eqn.\ (2.13) of \cite{P81}.

\section{Simulations of the Outer Edge of Saturn's B Ring}
\label{B ring}

The semimajor axis of the outer edge of Saturn's B ring is
$a_{\mbox{\scriptsize edge}}=117568\pm4$ km,
and that edge lies $\Delta a_2=12\pm4$ km 
exterior to Mimas' $m=2$
inner Lindblad resonance (ILR) (\citealt{SP10}, hereafter SP10). Evidently
Mimas' $m=2$ ILR is responsible for confining the B ring and preventing
it from viscously diffusing outwards and into the Cassini Division.
Mimas' $m=2$ ILR excites a forced disturbance at the ring-edge whose
radius--longitude relationship $r(\theta)$ is expected to have the form
$r(\theta, t) = a_{\mbox{\scriptsize edge}} - R_m\cos m(\theta - \tilde{\omega}_m)$
where $R_m$ is the epicyclic amplitude of the mode whose
azimuthal wavenumber is $m$ and whose orientation at time $t$
is given by the angle $\tilde{\omega}_m(t)$.
This forced disturbance is expected to corotate with Mimas' longitude, and such a
pattern would have a pattern speed $\dot{\tilde{\omega}}_m=d\tilde{\omega}_m/dt$ 
that satisfies $\dot{\tilde{\omega}}_m=\Omega_s$ where $\Omega_s$ 
is satellite Mimas' angular velocity.

SP10 have analyzed the many images of the B ring's edge that have been collected
by the Cassini spacecraft, and they show that this ring-edge does indeed
have a forced $m=2$ shape that corotates with Mimas as expected. But they also show
that the B ring's edge has an additional 
{\em free} $m=2$ pattern that rotates slightly faster
than the forced pattern. SP10 also detect two additional modes,
a slowly rotating $m=1$ pattern as well as a rapidly rotating $m=3$ pattern.
These findings are confirmed by stellar occulation observations
of the B ring's outer edge that also detect additional lower-amplitude
$m=4$ and $m=5$ modes \citep{NFH12}.

The following will use the N-body model to investigate the higher amplitude 
$m=1,2$, and 3 modes seen at the B ring's edge. But keep in mind that only the $m=2$
forced pattern has a known driver, namely, Mimas' $m=2$ ILR, while the nature of the
perturbation that launched the other three free modes in the B ring is quite unknown.
So to study the B ring's behavior when those free modes are present, an
admittedly ad hoc method is used. Specifically, the simulated ring particles'
initial conditions are constructed in a way that plants a free $m=1,2$, or 3
pattern at the simulated ring's edge at time $t=0$. The N-body integrator then advances
the system over time, which then reveals
how those free patterns evolve over time. And to elucidate those findings
most simply, the following subsections first 
consider the B ring's $m=1$, 2, and 3 patterns in isolation.

All simulations use a timestep $\Delta t=0.2/2\pi=0.0318$ orbit periods,
so there are 31.4 timesteps per orbit of the simulated B ring, and nearly all
simulations use oblateness $J_2=0.01629071$, which is the same value we used
in previous work \citep{HSP09}.

And lastly, these simulations also zero the viscous acceleration 
that is exerted at the simulated B
ring's innermost and outermost streamlines, to prevent them from drifting radially
due to the ring's viscous torque. This is in fact appropriate for the simulation's innermost
streamline, since in reality the viscous torque from the
unmodeled part of the B ring should deliver to the inner streamline a constant
angular momentum flux $F$ that it then communicates to the adjacent streamline, 
so the viscous acceleration $A_{\nu,\theta}\propto \partial F/\partial r$
at the simulation's inner edge really should be zero.
But zeroing the viscous acceleration of
outer streamline might seem like a slight-of-hand since it should be
$A_{\nu,\theta}=F/\lambda r$ according Section \ref{viscosity}.
But setting $A_{\nu,\theta}=0$ is done because, if not, then the outermost
streamline will slowly but steadily drifts radially outwards past Mimas' $m=2$ ILR, 
which also causes that streamline's forced eccentricity to slowly and steadily grow
as the streamline migrates.
This happens because the model does not settle into a balance where
the ring's positive viscous torque on its outermost streamline is opposed
by a negative torque exerted by the satellite's gravity. We also note that
the semi-analytic model of this resonant ring-edge, which is described in
\cite{HSP09}, also had the
same difficulty in finding a torque balance. So to sidestep this difficulty,
this model zeros the viscous acceleration at the outermost streamline,
which keeps its semimajor axis static as if it were in the expected torque
balance. This then allows us to compare simulations to the B ring's forced
$m=2$ pattern to that measured by the Cassini spacecraft. The validity of
this approximation is also assessed below in Section \ref{force}.

\subsection{the forced and free \boldmath{$m=2$} patterns}
\label{m=2}

SP10 detect a forced $m=2$ pattern at the B ring's outer edge
that has an epicyclic amplitude $R_2=34.6\pm0.4$ km,
and that forced pattern corotates with the satellite Mimas. They also detect a free
pattern whose epicyclic amplitude is $2.7$ km larger, so the forced and free 
patterns are nearly equal in amplitude, and the free pattern
rotates slightly faster than the forced pattern
by $\Delta\dot{\tilde{\omega}}_2=0.0896\pm0.0007$ degrees/day (SP10).
The radius-longitude relationship for a ring-edge that experiences
these two modes can be written
\begin{eqnarray}
    \label{dr_2}
    r(\theta, t) &=& a - R_2\cos m(\theta - \theta_s) - 
        \tilde{R}_2\cos m(\theta - \tilde{\omega}_2)
\end{eqnarray}
where $R_2$ is the epicyclic amplitude of the forced pattern that corotates with
Mimas whose longitude is $\theta_s(t)$ at time $t$, and $\tilde{R}_2$
is the epicyclic amplitude of the free pattern with
$\tilde{\omega}_2(t)$ being the free pattern's longitude.

The N-body integrator
{\tt epi\_int} is used to simulate the forced and free $m=2$ patterns that are seen 
at the outer edge of the B ring, for simulated rings having a variety of initial surface
densities $\sigma_0$. These simulations use $N_r=130$ streamlines that are
distributed uniformly in the radial direction with spacings $\Delta a=5.13$ km,
so the radial width of the simulated portion of the B ring is 
$w=(N_r - 1)\Delta a=662$ km. Each streamline is populated with $N_\theta=50$
particles that are initially distributed uniformly in longitude $\theta$ and in
circular coplanar orbits.
These simulations use a total of $N_rN_\theta=6500$ particles, which
is more than sufficient to resolve the $m=2$ patterns seen here. These systems
are evolved for $t=41.5$ years, which corresponds to $3.2\times10^4$ orbits,
and is sufficient time to see the simulation's slightly faster
free $m=2$ pattern lap the forced
$m=2$ pattern several times. The execution time for these high resolution,
publication-quality simulations is 1.5 days on a desktop PC, but sufficiently
useful preliminary results from lower-resolution simulations
can be obtained in just a few hours.

The B ring's viscosity is unknown, so these 
simulations will employ a value for the
kinematic shear viscosity $\nu_s$
and bulk viscosity $\nu_b$ that are typical of Saturn's A ring,
$\nu_s=\nu_b=100$ cm$^2$/sec \citep{TBN07}. 
The simulated particles' dispersion velocity $c$ is also chosen so
that the ring's gravitational stability parameter $Q=c\kappa/\pi G\sigma_0=2$,
since Saturn's main rings likely have $1\lesssim Q\lesssim2$ \citep{S95}. Mimas'
mass is $m_s=6.5994\times10^{-8}$ Saturn masses, and its semimajor axis $a_s$ is
chosen so that its $m=2$
inner Lindblad resonance lies $\Delta a_{\mbox{\scriptsize res}}=12.2$ km 
interior to the simulated B ring's outer edge. This model only accounts for
the $J_2=0.01629071$ part of Saturn's oblateness, so the constraint
on the resonance location puts the simulated Mimas at $a_s=185577.0$ km, which is 38 km
exterior to its real position.

\begin{figure}[th]
    \epsscale{0.9}
    \plotone{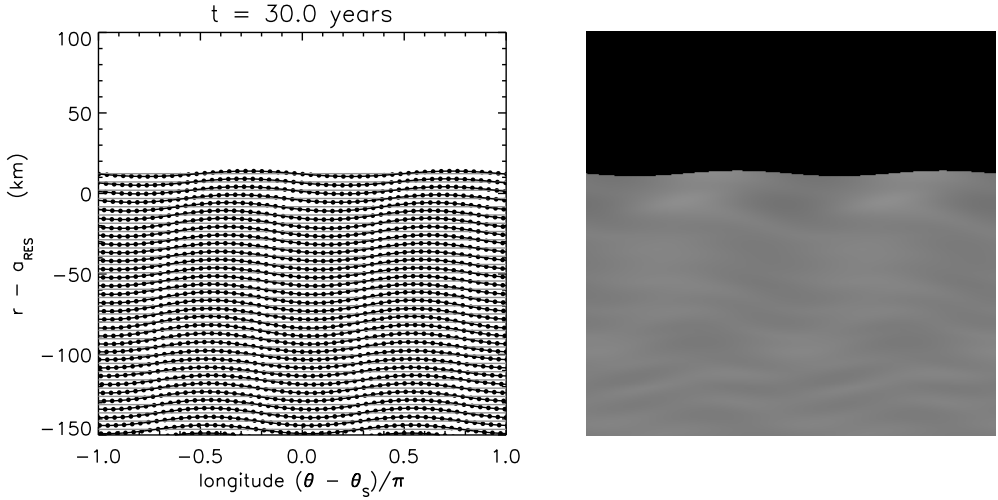}
    \figcaption{
        \label{m=2_fig}
        Five snapshots of a simulated B ring that is perturbed by Mimas' $m=2$ ILR. The simulated
        ring has an undisturbed surface density $\sigma_0=280$ gm/cm$^2$, with other model
        details provided in Section \ref{m=2}.
        Black curves show each distorted streamline via a radius versus longitude plot,
        with the streamline's radial displacement measured along the vertical axis and
        longitude measured along the horizontal axis. 
        Note that the simulated ring extends
        inwards another 420 km beyond that shown here.
        All distances are measured relative to the resonance radius $a_{res}$ 
        and all longitudes are measured relative to satellite Mimas' longitude $\theta_s$. 
        Dots indicate the locations of all particles, and gray lines 
        indicate their semimajor axes.
        The grayscale map shows the fractional
        variations in the ring's surface density $\sigma/\sigma_0$ scaled
        so that gray corresponds
        to an undisturbed region having $\sigma=\sigma_0$, black for regions where there
        is no ring matter, and white saturating in regions where the ring is overdense
        by at least two, $\sigma\ge2\sigma_0$. Keep in mind that the 
        particles sample the ring's
        surface density across an irregular grid, so to generate these grayscale maps,
        splines are first fit to each streamline
        so that each are resampled along a regularly spaced
        grid in longitude $\theta$. Then another set of splines
        are fit along each longitude to determine the radial distance $r(\theta)$ of
        each streamline along direction $\theta$, which then allows the ring's surface
        density $\sigma(r,\theta)$ to be determined along a grid that is uniformly
        sampled in $r,\theta$.\vspace*{-12ex}
    }
\end{figure}

Starting the ring particles in circular orbits provides an
easy way to plant equal-amplitude free and forced $m=2$ patterns in the ring.
This creates a free $m=2$ pattern
that  at time $t=0$  nulls perfectly the forced $m=2$ pattern due to Mimas.
However the free pattern rotates slightly
faster than the forced pattern, so the ring's epicyclic amplitude
varies between near zero and $\sim2R_2$ as the rotating patterns interfere
constructively or destructively over time. This behavior 
is illustrated in Fig.\ \ref{m=2_fig} which shows
results from a simulation of a B ring whose undisturbed surface density is
$\sigma_0=280$ gm/cm$^2$. The wire 
diagrams show the ring's streamlines via radius versus
longitude plots, with dots indicating individual particles, and
the adjacent grayscale map shows the ring's surface density at that instant.
Figure \ref{m=2_fig} shows snapshots of the system at five distinct times that 
span one cycle of the ring's circulation: at time $t=26.4$ yr
when the ring's outermost streamline is nearly circular due to the 
forced and free patterns being out of phase by nearly $180^\circ/m=90^\circ$ and interfering
destructively, to time $t=28.2$ yr when the forced and free patterns are in phase and
interfere constructively, to nearly circular again at time $t=30.0$ yr.

\pagebreak
\begin{figure}[th]
    \epsscale{0.9}
    \plotone{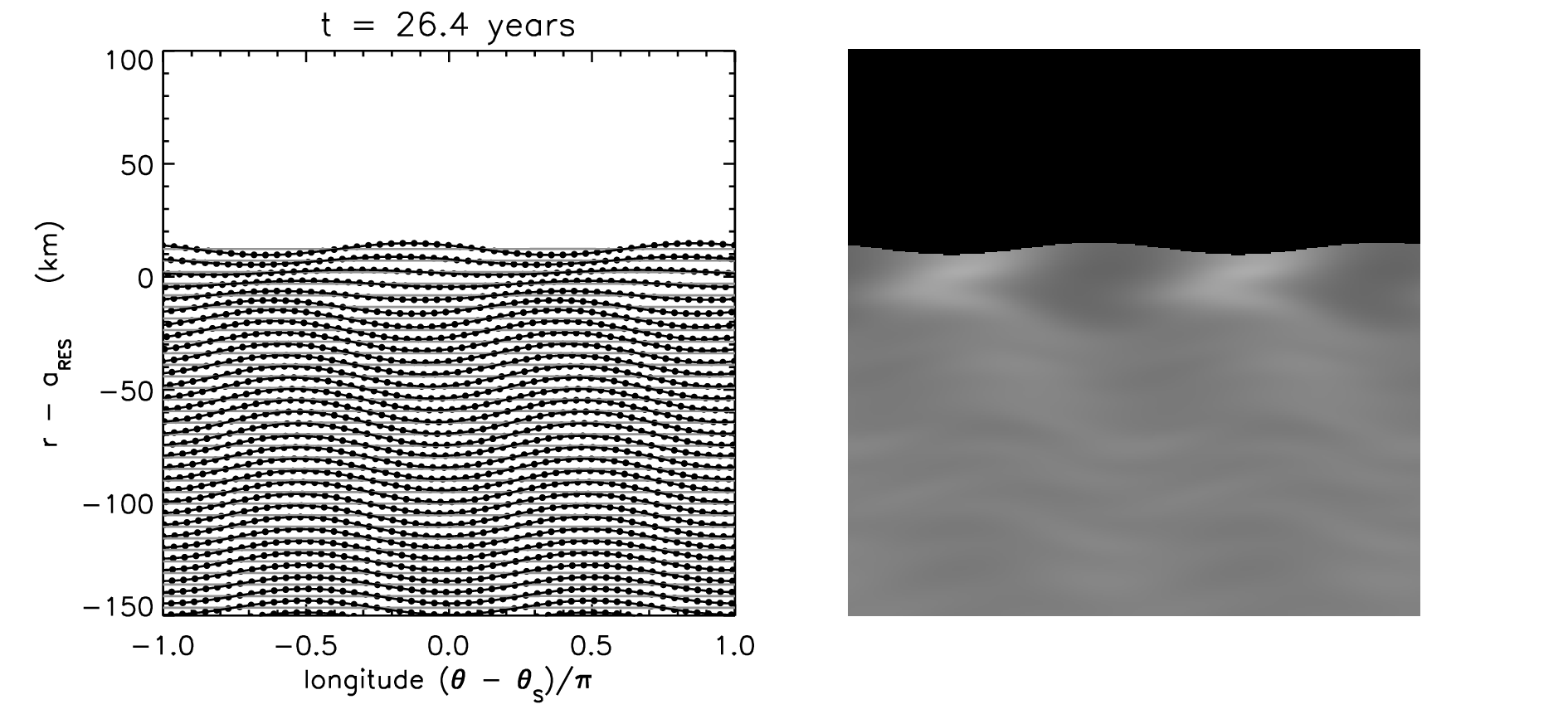}\vspace*{6ex}
    \plotone{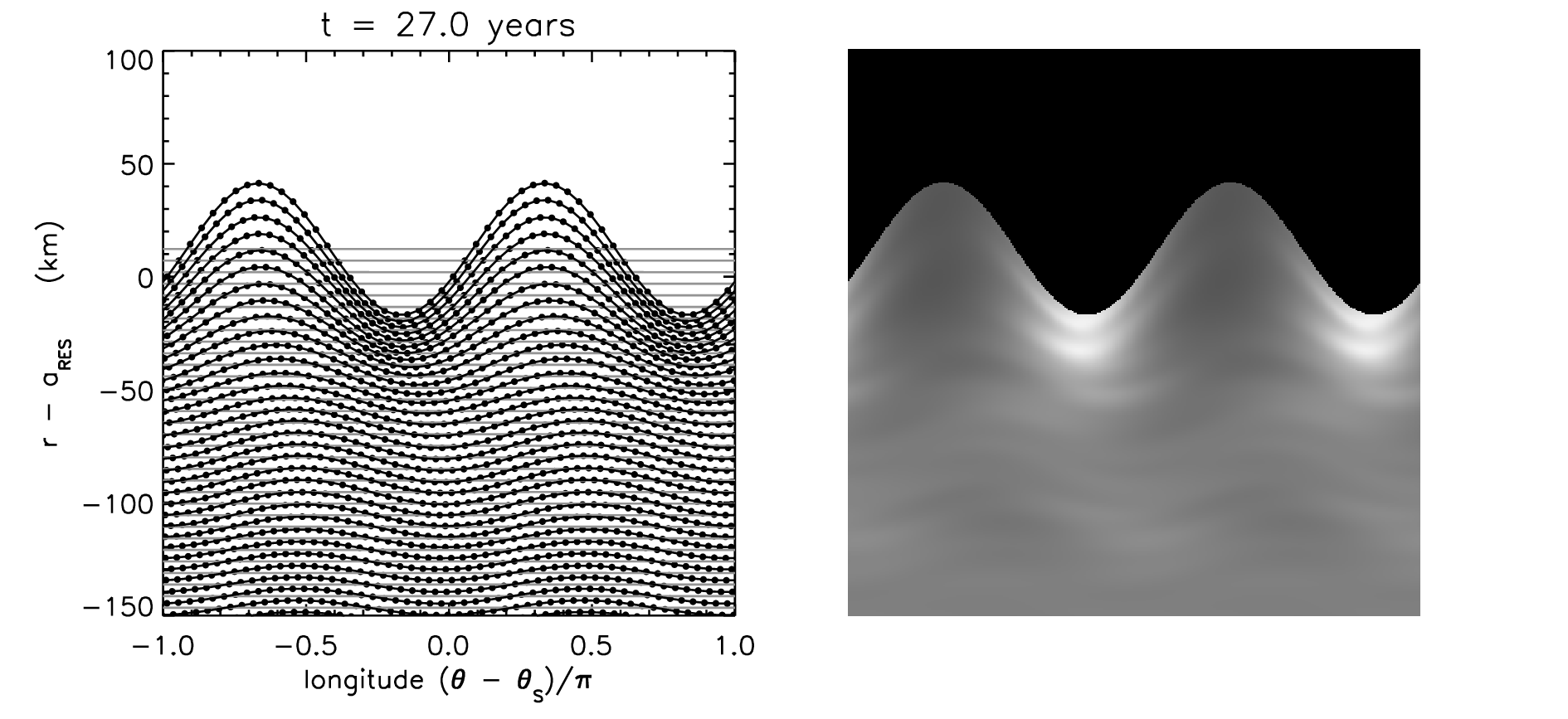}
\end{figure}

\pagebreak
\begin{figure}[th]
    \epsscale{0.9}
    \plotone{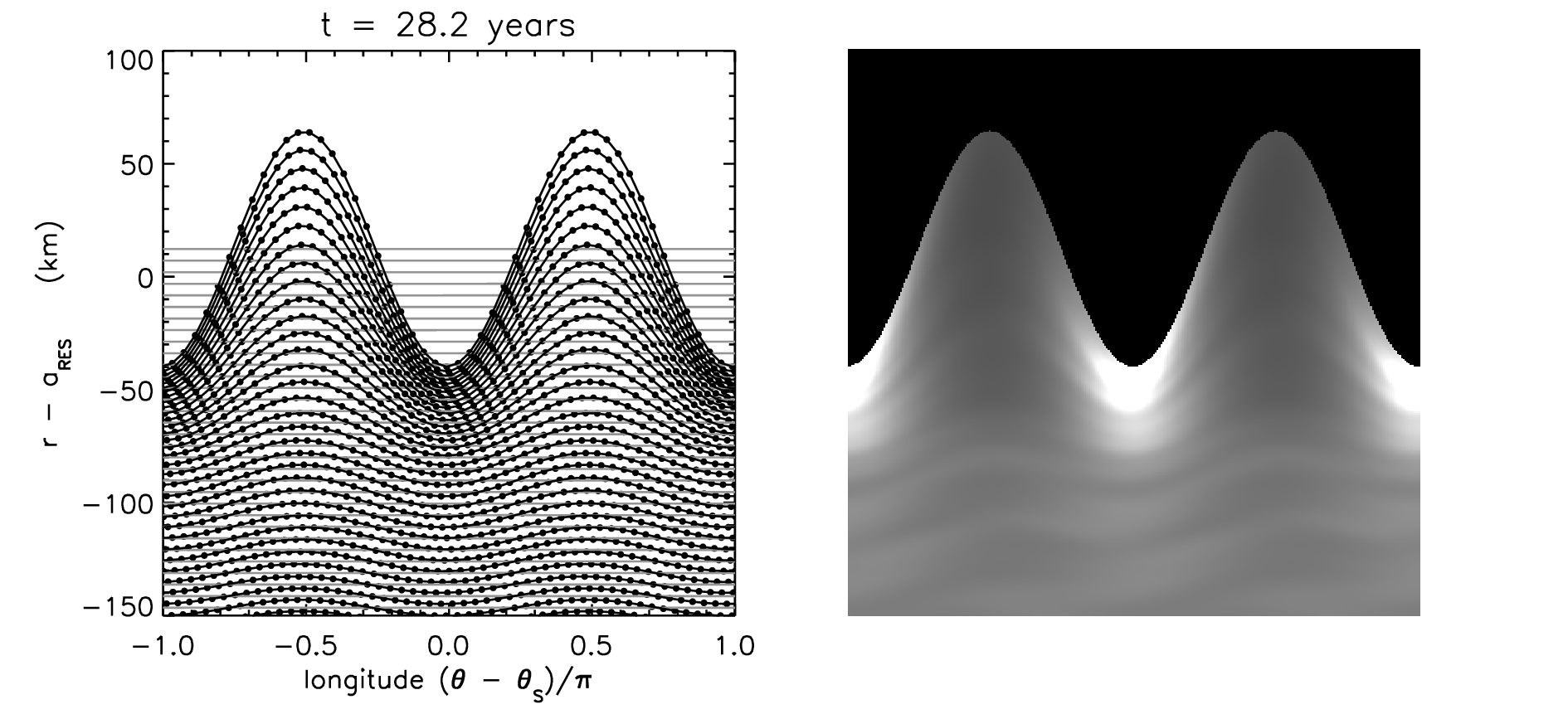}\vspace*{6ex}
    \plotone{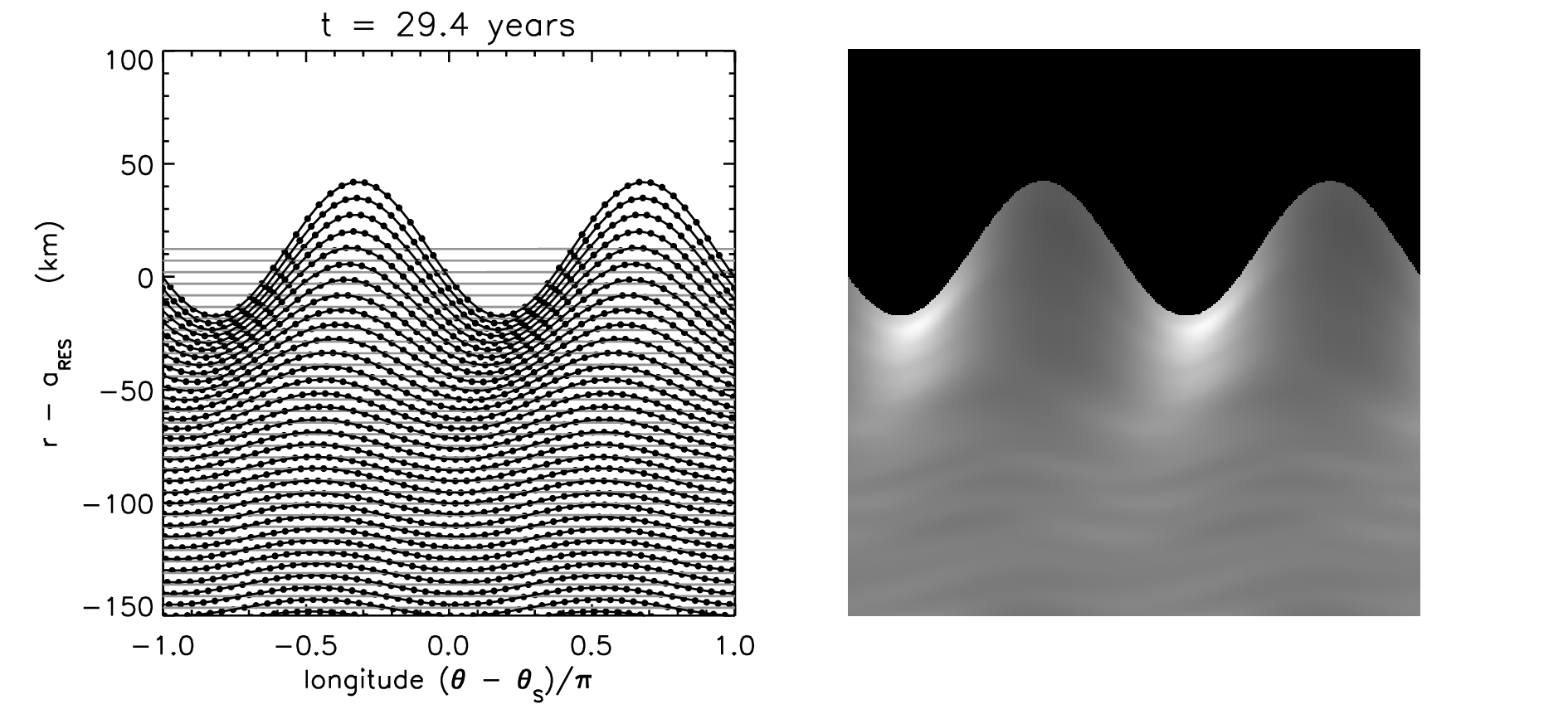}
\end{figure}

\pagebreak
\pagebreak

The circulation cycle seen in Fig.\ \ref{m=2_fig} repeats for the duration of the integration,
which spans about 10 cycles. The gray lines in Fig.\ \ref{m=2_fig} show the semimajor
axes $a$ of all particles on each streamline; 
note that all particles on a given streamline preserve
a common semimajor axes, and this is also true of their eccentricities $e$. 
In the simulations shown
here, the two orbit elements $a$ and $e$ do not vary with the particle's longitude $\theta$.
This however is distinct from the particles' angular orbit elements $M$ and $\tilde{\omega}$,
which do vary linearly with longitude $\theta$ along each streamline. 
Recall that the {\tt epi\_int}  code does not in any way force or require
particles to inhabit a given streamline. The streamline concept is only used when calculating
the forces that all of the ring's streamlines exert on each particle, which
the symplectic integrator then uses to advance these particles forwards in time.
Although a particle's $e$ and $a$ are in principle 
free to drift away from that of the other streamline-members,
that does not happen in the simulations shown here; evidently the particles'
$a$ and $e$ evolve slowly in the orbit-averaged sense, with that time-averaged
evolution being independent of longitude $\theta$. This accounts for
why all particles on the same streamline have the same evolution in $a$ and $e$.
This time-averaged evolution is also a standard assumption that is routinely
invoked in analytic models of planetary rings (see {\it cf}.\ \citealt{GT79, BGT86, HSP09}),
and the simulations shown here confirm the validity of that assumption.

\begin{figure}[th]
\epsscale{0.86}
\plotone{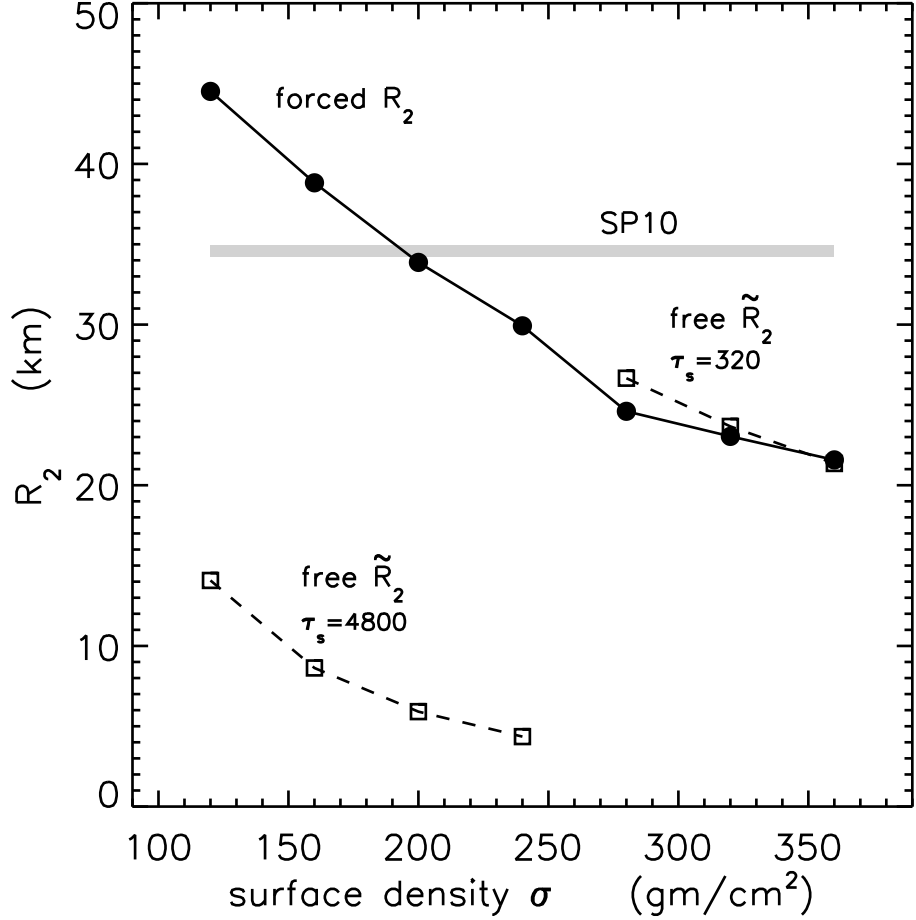}
\figcaption{
    \label{R2_fig}
    Solid curve is the
    epicyclic amplitude $R_2$ for the $m=2$ pattern forced by Mimas, plotted
    versus ring surface density $\sigma_0$
    for the B ring simulations described in Section \ref{m=2}. Dashed curve gives the simulated
    ring's free epicyclic amplitude $\tilde{R}_2$. Grey bar is the ring's
    observed forced $m=2$ epicyclic amplitude, from SP10; the bar's vertical extent,
    $\pm0.35$ km, spans the the uncertainty in the observed $R_2$. In simulations
    with $\sigma_0\le240$ gm/cm$^2$,
    Mimas is grown to its current mass over
    timescales $\tau_s=4800$ ring orbits, and simulations with $\sigma_0\ge280$ gm/cm$^2$
    have $\tau_s=320$ orbits.\vspace*{-10ex}
}
\end{figure}

A suite of seven B ring simulations is performed for rings whose
undisturbed surface densities range over $120\le\sigma_0\le360$ gm/cm$^2$. Results 
are summarized in Fig.\ \ref{R2_fig} which shows the forced epicyclic amplitude $R_2$
(solid curve) and the free epicyclic amplitude $\tilde{R}_2$ (dashed curve) from each simulation.
These amplitudes are obtained by fitting Eqn.\ (\ref{dr_2}) to the simulated B
ring's outermost streamline assuming that the free pattern there rotates at a constant rate,
$\tilde{\omega}_2(t) = \tilde{\omega}_0 + \dot{\tilde{\omega}}_2t$
where $\tilde{\omega}_0$ is the free pattern's angular offset at time $t=0$
and $\dot{\tilde{\omega}}_2$ is the free mode's pattern speed. 
Equation (\ref{dr_2}) provides 
an excellent representation of the ring-edge's behavior over time, and that equation
has four parameters $R_2, \tilde{R}_2, \tilde{\omega}_0$, and 
$\dot{\tilde{\omega}}_2$ that are determined by least squares fitting.
The observed epicyclic amplitude of the B ring's forced $m=2$ component is
$R_2=34.6\pm0.4$ km (SP10), and the gray bar in Fig.\ \ref{R2_fig} indicates
that the outer edge of the B ring has a surface density of about
$\sigma_0=195$ gm/cm$^2$. And if we naively assume that the ring's surface density is
everywhere the same, then its total mass of Saturn's B ring is 
about $90\%$ of Mimas' mass.

Figure \ref{R2_fig} also shows that the amplitude of the forced pattern $R_2$
gets larger for rings that have a smaller surface density $\sigma_0$, due to the ring's
lower inertia, with the forced response varying roughly as
$R_2\propto\sigma_0^{-0.67}$.
This also makes lighter rings more difficult to simulate, 
because their larger epicyclic amplitudes also causes the ring's streamlines
to get more bunched up at periapse. For instance in the $\sigma_0=280$ gm/cm$^2$ simulation
of Fig.\ \ref{m=2_fig}, the ring's edge at longitudes
$\theta=\theta_s$ and $\theta=\theta_s\pm\pi$  are overdense by a factor
of 3 at time $t=28.2$ yr, which is when the force and free patterns add constructively.
Streamline bunching in lighter rings is even more extreme, which is also more problematic,
because streamlines that are too compressed can at times cross in these overdense sites, and
the simulated ring's subsequent evolution becomes unreliable.

\pagebreak

To avoid the streamline crossing that occurs in simulations of lower surface density, 
the model also grows the mass of Mimas exponentially 
over the timescale $\tau_s$ that takes
values of $0.41\le \tau_s\le6.2$ years, with faster satellite growth ($\tau_s=0.41$ yrs
or 320 B ring orbits) occurring
in simulations of a heavy B ring having $\sigma_0\ge280$ gm/cm$^2$
and slower growth ($\tau_s=6.2$ yrs or 4800 B ring orbits)
for the lighter $\sigma_0\le240$ gm/cm$^2$ ring simulations.
The satellite growth timescale $\tau_s$ controls the amplitude of the free pattern 
$\tilde{R}_2$, with the ring having a smaller free epicyclic amplitude
$\tilde{R}_2$ when $\tau_s$ is larger; see the dashed curve in Fig.\ \ref{R2_fig}. 
Indeed, when the satellite grows over a timescale
$\tau_s\gg6.2$ yrs 
({\it i.e.} $\tau_s\gg4800$ orbits), the ring responds adiabatically to forcing
by the slowly growing Mimas, and shows only a forced $m=2$ pattern that corotates 
with Mimas, with the free $m=2$ pattern having a negligible amplitude.
Consequently, only the  $\sigma_0=280, 320$, and $360$ gm/cm$^2$ simulations
in Fig.\ \ref{R2_fig} are faithful in their attempt to reproduce a B ring whose
free epicyclic amplitude $\tilde{R}_2$ is slightly larger than the forced amplitude $R_2$. 
However the lower-surface density simulations have free patterns whose amplitudes
are smaller than the forced patterns, and these simulated rings have outer edges
whose longitude of periapse librate about Mimas' longitude, rather than circulate.

Also of interest here is the so-called radial depth of the $m=2$ disturbance, 
$\Delta a_{e/10}$, which is defined
as the semimajor axis separation between the ring's
outer edge and the streamline whose mean eccentricity is one-tenth that of
the edge's eccentricity. For these $m=2$ simulations the radial depth is
$\Delta a_{e/10}=154$km, so the radial width of the simulated part of the ring is
$w=4.3\Delta a_{e/10}$.

\subsubsection{sensitivity to resonance location and other factors}
\label{edge_sensitivity}

The surface density $\sigma_0$ that is inferred from the amplitude of the ring's forced
motion $R_2$ is very sensitive to the uncertainty in the ring's
semimajor axis, which is $\delta a_{\mbox{\scriptsize edge}}$. For example,
when the B ring is simulated again but with its outer edge instead 
extending further out by $\delta a_{\mbox{\scriptsize edge}}=4$ km, those simulations
show that the ring's forced amplitude $R_2$ is larger by about 6 km,
which requires increasing $\sigma_0$ by $\delta\sigma_0=60$ gm/cm$^2$ so that the simulated
$R_2$ is in agreement with the observed value. Similarly, when the simulated
ring's edge is moved inwards by $\delta a_{\mbox{\scriptsize edge}}=4$ km,
the forced amplitude $R_2$ is smaller and the ring's surface density 
$\sigma_0$ must be decreased by $\delta\sigma_0$ to compensate. So the surface density
of the B ring-edge is $\sigma_0=195\pm60$ gm/cm$^2$, and this value represents
the mean surface density of outer $\Delta a_{e/10}\simeq150$km that is most
strongly disturbed by Mimas' $m=2$ resonance. These results are also in excellent
agreement with the semi-analytic model of \cite{HSP09}, which calculated only
the ring's forced motion.

However these results are very insensitive to the model's other main unknown,
the ring's viscosity $\nu$. For instance, when we re-run the
$\sigma_0=200$ gm/cm$^2$ simulation with the ring's shear and bulk
viscosities increases as well as decreased by a factor of 10, 
we obtain the same forced response $R_2$. So these findings are 
insensitive to range of ring viscosities considered here,
$10<\nu<1000$ cm$^2$/sec.

\subsubsection{free $m=2$ pattern}
\label{free m=2}

\begin{figure}[th]
\epsscale{0.86}
\plotone{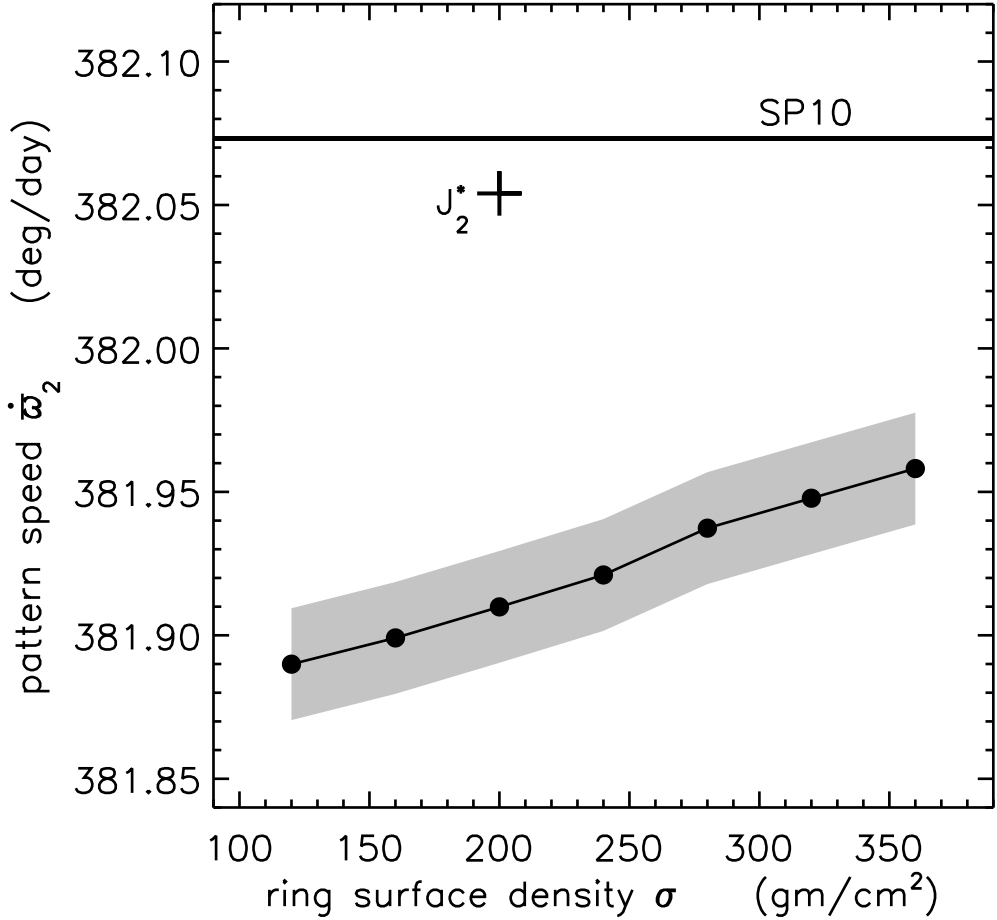}
\figcaption{
    \label{m2_fig}
    Curve with black dots is the pattern speed $\dot{\tilde{\omega}}_2$
    for the simulated B ring's free $m=2$ pattern.
    The vertical extent of the gray band indicates how these simulated
    results would change if the ring-edge's semimajor axis was altered
    by its observed uncertainty $\delta a_{\mbox{\scriptsize edge}}=\pm4$ km.
    The horizontal line is the B ring's
    observed free $m=2$ pattern speed $\dot{\tilde{\omega}}_2=382.0731\pm0.0007$
    deg/day, from SP10, with its small uncertainty indicated by the line's
    thickness. The cross shows how the free pattern speed in the
    $\sigma_0=200$ gm/cm$^2$ simulation changes when $J_2$ is boosted by
    factor $f^\star=1.0395013$ to $J_2^\star$.\vspace*{-15ex}
}
\end{figure}

The dotted curve in
Fig.\ \ref{m2_fig} shows the simulations' free $m=2$ 
pattern speeds $\dot{\tilde{\omega}}_2$, which is also sensitive to the
ring's undisturbed surface density $\sigma_0$.
The purpose of this subsection is to illustrate how a free normal mode
can also be used to determine the ring's surface density.
Although these result will not be as definitive as the value of $\sigma_0$
that was inferred from the ring's forced pattern, due to a greater sensitivity
to the observational uncertainties,
the following illustrates an alternate technique that in principle
can be used to infer the surface density of other rings, such as the
many narrow ringlets orbiting Saturn that also exhibit
free normal modes.

\pagebreak
\pagebreak

But first note the models' large discrepancy with the observed free $m=2$
pattern speed reported in SP10, which is the upper horizontal bar in Fig.\ \ref{m2_fig}.
This discrepancy is {\em not} due to the
$\delta a_{\mbox{\scriptsize edge}}=\pm4$km uncertainty in the
ring-edge's semimajor axis, which makes the simulated ring particles'
mean angular velocity uncertain by the fraction 
$\delta\Omega/\Omega=1.5\delta a_{\mbox{\scriptsize edge}}/a_{\mbox{\scriptsize edge}}
\simeq 0.005\%$. We find empirically
that the simulations' pattern speeds are also uncertain
by this fraction, so $\delta\dot{\tilde{\omega}}_2\simeq0.02$ deg/day,
which is the vertical extent of the gray band around the simulated data
in Fig.\ \ref{m2_fig}.

Rather, this discrepancy is indirectly due to the absence of the $J_4$
and higher terms from the N-body simulations. To demonstrate this, repeat the
$\sigma_0=200$ gm/cm$^2$ simulation with $J_2$ boosted slightly by factor
$f^\star=1.0395013$ so that the second zonal harmonic is 
$J_2^\star=f^\star J_2=0.016934294$.
This increases the simulated B ring-edge's angular velocity slightly to
$\Omega_{\mbox{\scriptsize edge}}=758.8824$ deg/day, which is in fact
the ring's true angular velocity at $a=a_{\mbox{\scriptsize edge}}$
when the higher order $J_4$ and $J_6$
terms are also accounted for\footnote{This mean angular velocity is
obtained using the physical constants given
in the 25 August 2011 Cassini SPICE kernel file: $Gm_0=37940585.47323534$ km$^3$/sec$^2$,
$J_2=0.016290787119$, $J_4=-0.000934741301$, and $J_6=0.000089240275$.}. 
And since Saturn's gravitational force there is
$a_{\mbox{\scriptsize edge}}\Omega_{\mbox{\scriptsize edge}}^2$, this means
that Saturn's gravity on the simulated particles
at $r=a_{\mbox{\scriptsize edge}}$ is in fact the true value.
Note that boosting $J_2$ to the slightly larger value $J_2^\star$ also requires
bringing the simulated Mimas
inwards and just interior to its true semimajor axis by 2km. Which speeds up
both the forced and free pattern speeds, and is why this simulation's 
free $m=2$ pattern speed $\dot{\tilde{\omega}}_2$, which is the 
cross in Fig.\ \ref{m2_fig}, is in better agreement with the observed
pattern speed. So the discrepancy between all the other simulated and observed
pattern speeds $\dot{\tilde{\omega}}_2$ is due to
those models' not accounting for the additional gravity that is due to the $J_4$ and higher
terms in Saturn's oblate figure. Compensating for the absence of those
oblateness effects requires altering the simulated satellite's orbits slightly, which in turn
alters the forced and free pattern speeds slightly.
But the following will show that these two patterns' {\em relative} speeds
are quite insensitive to the particular value of $J_2$
and the absence of the $J_4$ and higher terms.

\begin{figure}[th]
\epsscale{0.86}
\plotone{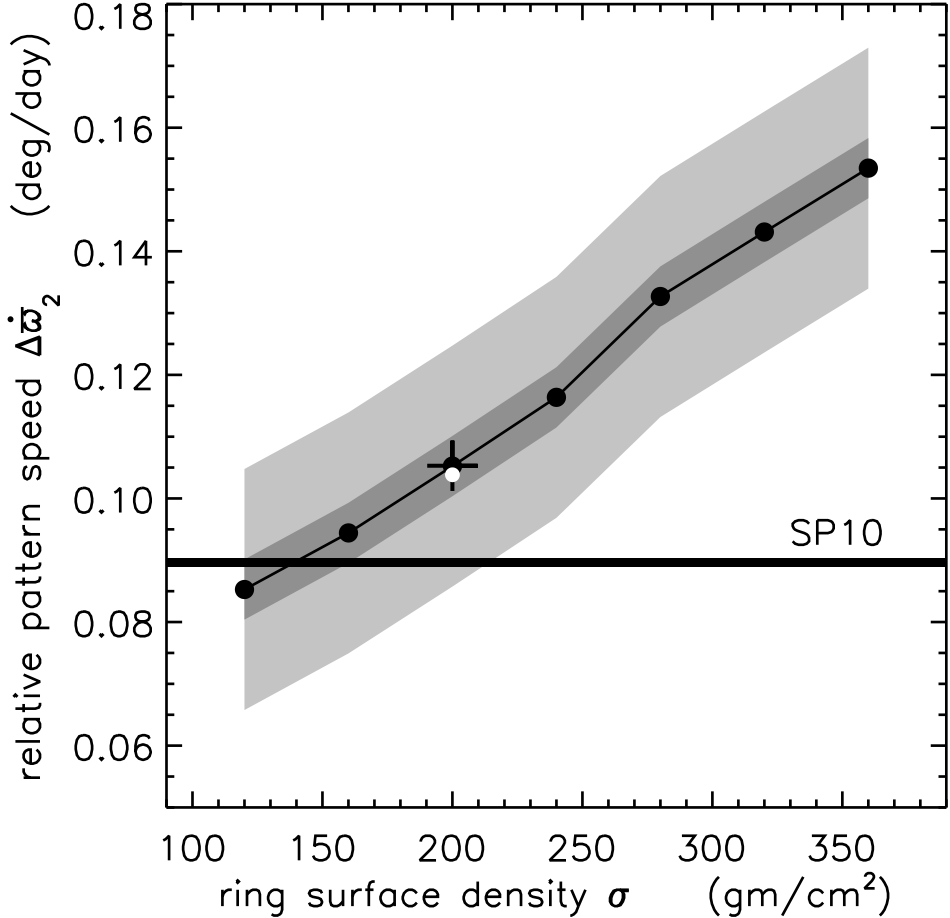}
\figcaption{
    \label{m2_rel_fig}
    Dots indicate the simulations' free $m=2$ pattern speed
    relative to the force pattern speed,
    $\Delta\dot{\tilde{\omega}}_2=\dot{\tilde{\omega}}_2-\Omega_{\mbox{\scriptsize sat}}$,
    with the light gray indicating the uncertainty due to the
    $\delta a_{\mbox{\scriptsize edge}}=\pm4$ km uncertainty in the 
    ring-edge's semimajor axis. The dark gray indicates what would result if 
    $\delta a_{\mbox{\scriptsize edge}}$ were instead $\pm1$ km.
    The horizontal line is the B ring's
    observed $m=2$ relative pattern speed from SP10, with
    the small uncertainty indicated by the line's
    thickness. The cross shows indicates that the free pattern's relative
    speed in the $\sigma_0=200$ gm/cm$^2$ simulation is unchanged
    when $J_2$ is boosted by
    factor $f^\star=1.0395013$ to $J_2^\star$, and the white dot indicates that the
    relative pattern speed when $J_2$ is instead set to zero.
    All simulations have Mimas' orbit configured so that its forced $m=2$ inner
    Lindblad resonance
    lies 12.2 km interior to the semimajor axis of the B ring's outer edge.\vspace*{-25ex}
}
\end{figure}

The best way to compare simulated to observed free $m=2$ patterns is to consider the
free $m=2$ pattern speed relative to the forced pattern speed, which is the
satellite's mean angular velocity $\Omega_{\mbox{\scriptsize sat}}$.
That frequency difference is 
$\Delta\dot{\tilde{\omega}}_2=\dot{\tilde{\omega}}_2-\Omega_{\mbox{\scriptsize sat}}$,
and is plotted versus ring surface density $\sigma_0$ in Fig.\ \ref{m2_rel_fig}.
Black dots are from the simulation and
the light gray band indicates the $\delta\dot{\tilde{\omega}}_2\simeq0.02$ deg/day
spread that results from the  $\delta a_{\mbox{\scriptsize edge}}=\pm4$ km 
uncertainty in the ring-edge's semimajor axis.
The relatively large uncertainty in $a_{\mbox{\scriptsize edge}}$
means that one can only conclude from Fig.\ \ref{m2_rel_fig}
that $\sigma_0\lesssim210$ gm/cm$^2$. If however the uncertainty in
$a_{\mbox{\scriptsize edge}}$ were instead $\delta a_{\mbox{\scriptsize edge}}=\pm1$ km,
then the uncertainty in $\Delta\dot{\tilde{\omega}}_2$ would be 4 times smaller
(darker gray band), which would have allowed us to determine the ring surface density
with a much smaller uncertainty of only $\pm20$ gm/cm$^2$.
The lesson here is that if one wishes to use
models of free patterns to infer $\sigma_0$ in, say, narrow ringlets,
then one will likely need to know the ring-edge's semimajor axis with a precision of
$\delta a_{\mbox{\scriptsize edge}}\simeq\pm1$ km.

\pagebreak

The cross in Fig.\ \ref{m2_rel_fig} indicates that the the free $m=2$ pattern
speed relative to the forced is unchanged when Saturn's oblateness is boosted
to $J_2^\star$. And to demonstrate that this kind of plot is rather insensitive
to oblateness effects, the white dot in Fig.\ \ref{m2_rel_fig} shows that these
relative pattern speeds change only very slightly even when $J_2$ is set to zero.

\begin{figure}[th]
\epsscale{0.86}
\plotone{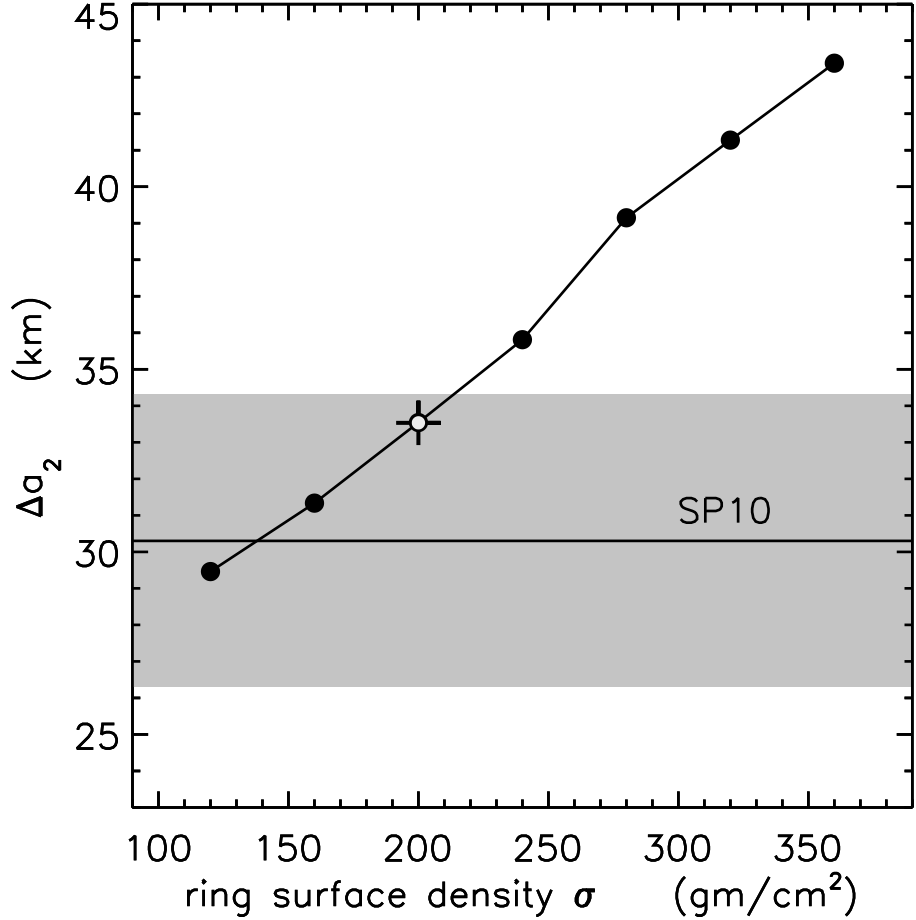}
\figcaption{
    \label{da2_fig}
    Eqn.\ (\ref{resonance_eqn}) is solved for the radius $\tilde{a}_2$ of the 
    inner Lindblad resonance that is associated with each of the
    simulated free $m=2$ modes
    whose pattern speeds $\Omega_{ps}=\dot{\tilde{\omega}}_m$ are
    shown in Fig.\ \ref{m2_fig}, with this Figure showing
    the relative distance $\Delta a_2 = a_{\mbox{\scriptsize edge}} - \tilde{a}_2$
    of the ILR from the semimajor axis of the simulated B ring's outer edge.
    The observed value is $\Delta a_2=30.3\pm4$ km (SP10) whose uncertainty
    is indicated by gray band. The cross and the white dot indicate that the results are
    unchanged when the oblateness
    parameter takes values $J_2^\star$ and zero.\vspace*{-15ex}
}
\end{figure}

Note though that there will be instances where there is no forced mode with which to
compare pattern speeds. In that case it will be convenient 
to convert the 
free pattern speed $\dot{\tilde{\omega}}_m=\Omega_{ps}$ into a  radius by solving the
Lindblad resonance criterion
\begin{eqnarray}
    \label{resonance_eqn}
    \kappa(r) &= \epsilon m[\Omega(r) - \Omega_{ps}]
\end{eqnarray}
for the resonance radius $r=a_m$,
where $\kappa(r)$ is the ring particles' epicyclic frequency (Eqn.\ \ref{kappa^2}),
and $\epsilon=+1 (-1)$ at an inner (outer) Lindblad resonance. So for the 
simulated B ring's free $m=2$ mode, Eqn.\ (\ref{resonance_eqn}) is solved for the
radius $r=\tilde{a}_2$ of the $\epsilon=+1$
inner Lindblad resonance that is associated with this mode. That quantity
is to be compared to a nearby reference distance, which in this case would be the 
semimajor axis of the B ring's outer edge  $a_{\mbox{\scriptsize edge}}$.
Results are shown in Fig.\ \ref{da2_fig}, which shows the simulations'
distance from the B ring's outer edge to the free $m=2$ pattern's ILR ,
$\Delta a_2 = a_{\mbox{\scriptsize edge}} - \tilde{a}_2$, plotted versus ring 
surface density $\sigma_0$. Heavier rings have a faster pattern speeds
(Fig.\ \ref{m2_fig}--\ref{m2_rel_fig}), and so the pattern's resonance
resides at a higher orbital frequency $\Omega(r)$ and thus must
lie further inwards of the ring's outer edge
in order to satisfy the resonance condition, Eqn.\ (\ref{resonance_eqn}).
Figure \ref{da2_fig} has the same information content as Fig.\ \ref{m2_rel_fig},
which is why it also tells us that the B ring's outer edge has 
$\sigma_0\lesssim210$ gm/cm$^2$. However a plot like Fig.\ \ref{da2_fig} will also
provide the best way to interpret the B ring's free $m=3$ mode, 
which is examined below in subsection \ref{m=3}.

\begin{figure}[th]
\epsscale{0.86}
\plotone{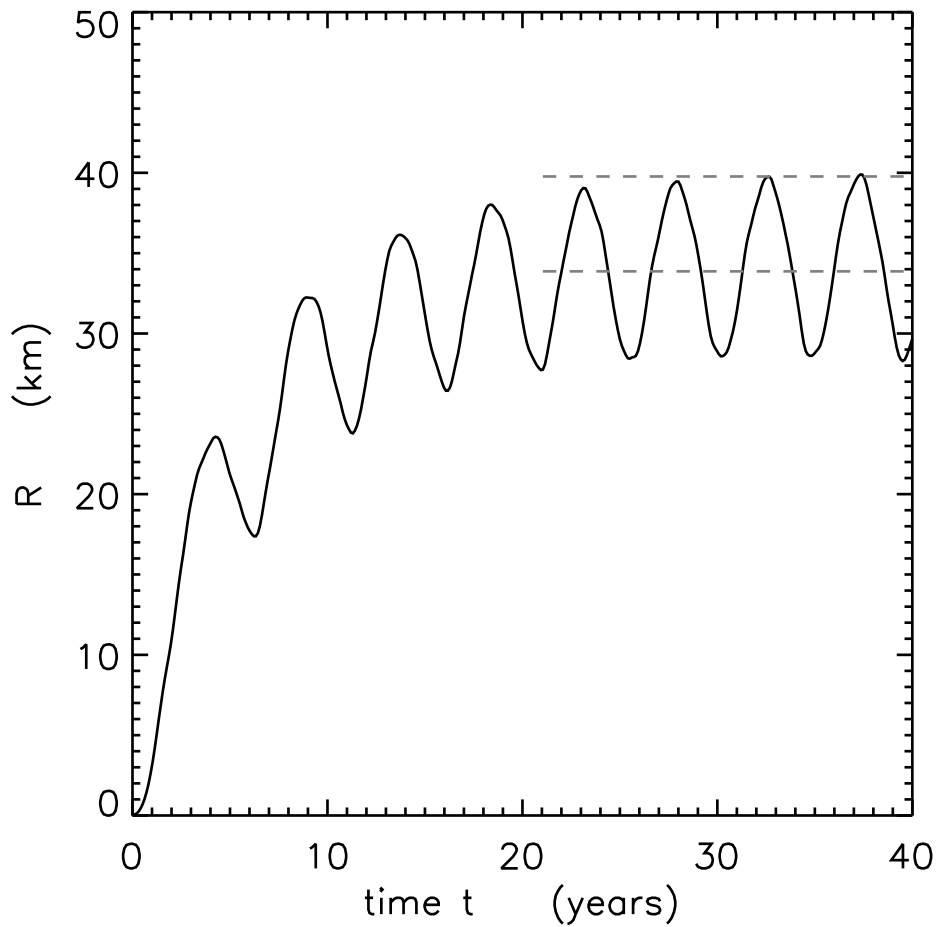}
\figcaption{
    \label{R_epi_fig}
    Epicyclic amplitude $R=|r-a|$ of the simulated B ring's outer edge over time $t$, from the
    $\sigma_0=200$ gm/cm$^2$ simulation shown in Figs.\ \ref{R2_fig}--\ref{da2_fig},
    with Mimas's mass grown exponentially over a $\tau_s=6.2$ year timescale.
    Lower dashed line is the amplitude of the ring's forced response $R_2$ due to Mimas'
    resonant perturbation, and the upper dash is the sum of the amplitudes of
    the ring's forced + free response $R_2 + \tilde{R}_2$, obtained by fitting
    Eqn.\ (\ref{dr_2}) to the curve at times $t>20$ years.\vspace*{-15ex}
}
\end{figure}

Lastly, note that the free $m=2$ patterns seen in these simulations persist for
$3\times10^4$ orbits or 40 years without any sign of damping, despite the ring's
viscosity $\nu=100$ cm$^2$/sec. This is illustrated in Fig.\ \ref{R_epi_fig}, which
plots the ring-edge's epicyclic amplitude over time for the nominal
$\sigma_0=200$ gm/cm$^2$ simulation. Indeed we have also rerun this
simulation using a viscosity that is ten times larger and still saw no damping.
These experiments reveal a possibly surprising result, that a free pattern can persist at
a ring-edge for a considerable length of time,
likely hundreds of years or longer, and Section \ref{force}
will show that this longevity is due to the viscous forces being several orders
or magnitude weaker than the ring's other interval forces. So one
possible interpretation of the free modes seen at the B ring and
at other ring edges is that they are relics from past disturbances in Saturn's ring
that may have happened hundreds or more years ago. This possibility is discussed
further in Section \ref{impulse}.

\newpage
\pagebreak

\newpage
\pagebreak

\subsection{the free \boldmath{$m=3$} pattern}
\label{m=3}

The B ring's free $m=3$ mode has an epicyclic amplitude of $\tilde{R}_3=11.8\pm0.2$ km,
a pattern speed $\dot{\tilde{\omega}}_3=507.700\pm0.001$ deg/day,
and the inner Lindblad resonance associated with this pattern speed lies 
$\Delta a_3=24\pm4$ km interior to the B ring's outer edge (SP10).

\begin{figure}[th]
\epsscale{0.86}
\plotone{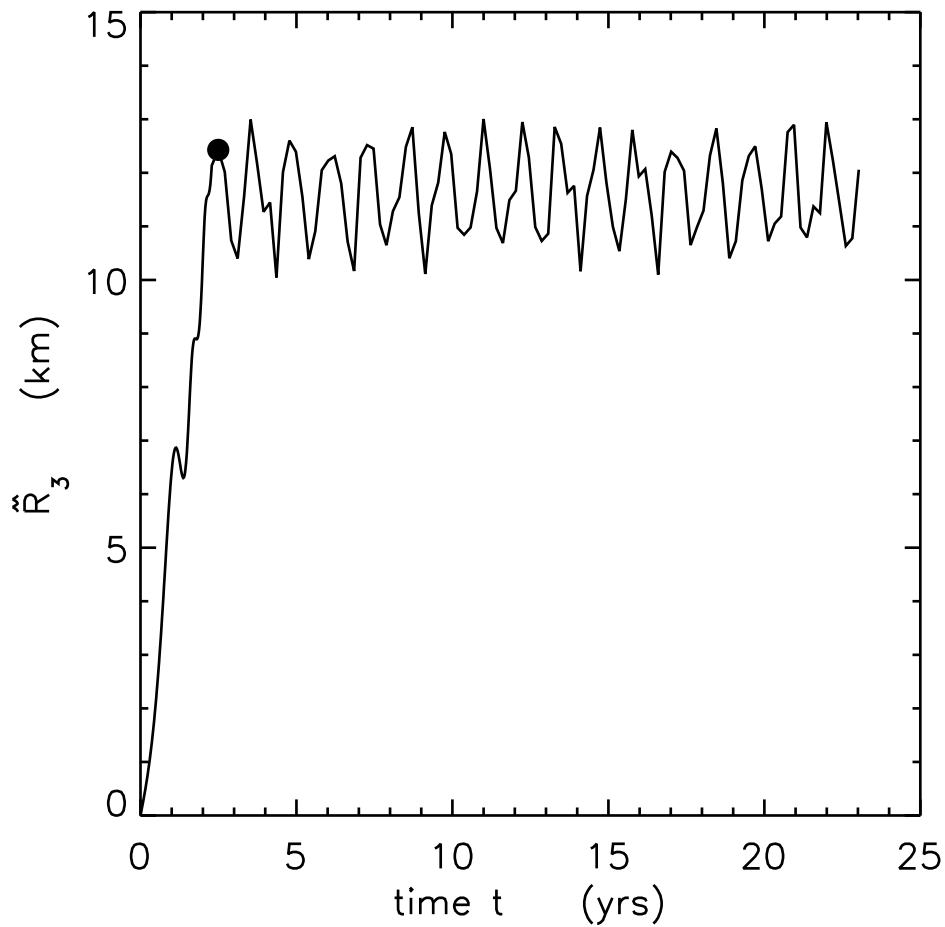}
\figcaption{
    \label{m3_fig}
    Epicyclic amplitude $\tilde{R}_3$ versus time $t$ for a simulated B ring 
    having surface density $\sigma_0=200$ gm/cm$^2$ that is perturbed
    until time $t=2.3$ yrs (black dot) by a satellite whose $m=3$ 
    inner Lindblad resonance lies 
    $\Delta a_{\mbox{\scriptsize res}}=24$ km interior to the ring's outer edge.
    The satellite's mass is Janus',
    and the dot indicates the time when that satellite is removed from the system,
    which converts this $m=3$ pattern into a unforced normal mode.\vspace*{-15ex}
}
\end{figure}

To excite a free $m=3$ pattern at the ring-edge,
place a fictitious satellite in an orbit that has an $m=3$
inner Lindblad resonance 
$\Delta a_3=24$ km interior to the ring's outer edge.
Noting that the satellite Janus happens to have an $m=3$
resonance in the vicinity,
about 2000 km inwards of the B ring's edge,
these simulations use a Janus-mass satellite
to perturb the simulated ring for about $1650$ orbits (about 2 years), which excites an
$m=3$ pattern at the ring's outer edge. 
The satellite is then removed from the system, which converts the pattern into
a free normal mode, and {\tt epi\_int} is then used to evolve the now unperturbed 
ring for another $1.8\times10^4$ orbits (about 23 years). Figure \ref{m3_fig}
plots the ring-edge's epicyclic amplitude, where it is shown
that the free mode persists at the B ring's outer edge, undamped over
time, despite the simulated ring's viscosity of $\nu=100$ cm$^2$/sec.

\begin{figure}[th]
\epsscale{0.86}
\plotone{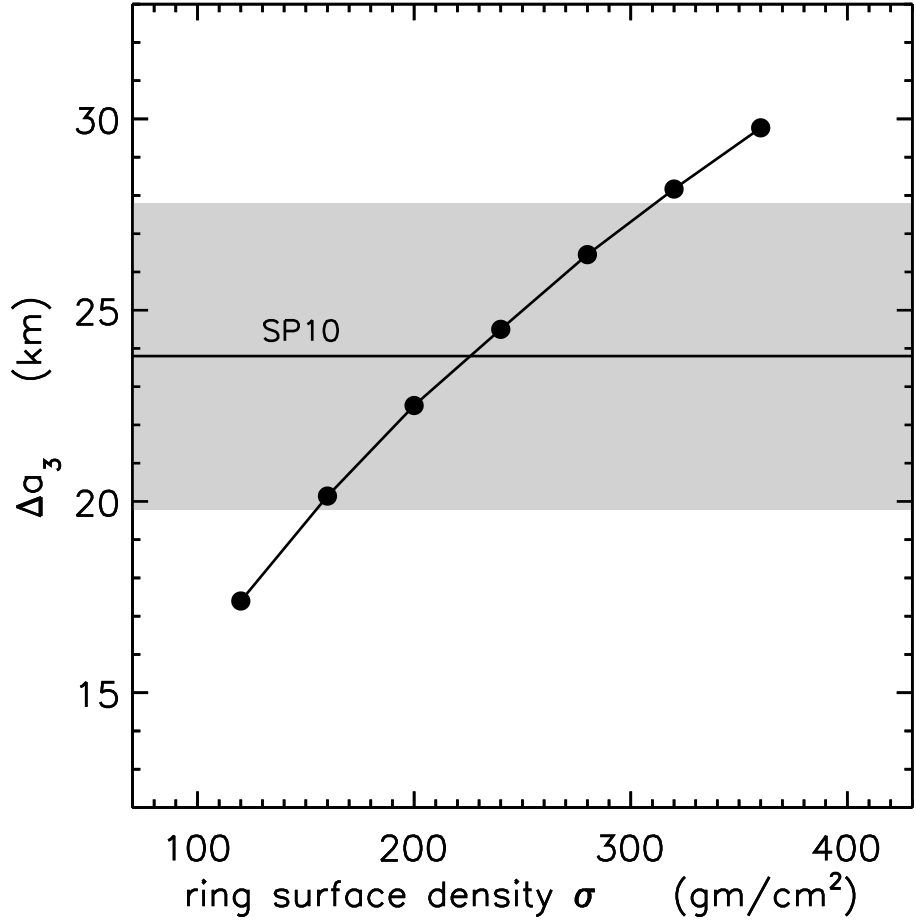}
\figcaption{
    \label{da3_fig}
    Distance $\Delta a_3 = a_{\mbox{\scriptsize edge}} - \tilde{a}_3$
    between the ring's outer edge and the
    inner Lindblad resonance associated with the free normal
    modes in the B ring simulations described in Section \ref{m=3},
    plotted versus ring surface density $\sigma_0$. The horizontal line
    is the observed distance with its uncertainty indicated
    by the gray band, from SP10. These simulations use $N_r=100$ streamlines
    with $N_\theta=60$ particles per and $N_rN_\theta=6000$ particles total.
    The streamlines' radial separation is $\Delta a=2.03$ km, and the total
    radial width of the simulated ring is $w=(N_r - 1)\Delta a=203$ km.\vspace*{-15ex}
}
\end{figure}

A suite of such B ring simulations is performed, with ring
surface densities $120\le\sigma_0\le360$ gm/cm$^2$ and all other parameters
identical to the nominal model of Section \ref{m=2} except where
noted in Fig.\ \ref{da3_fig} caption. The pattern speed
$\Omega_{ps}=\dot{\tilde{\omega}}_3$ of the $m=3$ normal mode
is then extracted from each simulation, with those speeds
again being slightly faster in the heavier rings.
Those pattern speeds are then inserted into Eqn.\ (\ref{resonance_eqn})
which is solved for the radius of the inner Lindblad resonance
$\tilde{a}_3$, each of which lies a distance 
$\Delta a_3 = a_{\mbox{\scriptsize edge}} - \tilde{a}_3$
inwards of the ring's outer edge, and those distances are plotted in
Fig.\ \ref{da3_fig} versus ring surface density $\sigma_0$.
The simulated distances $\Delta a_3$ are compared to the observed
edge-resonance distance reported in SP10, which indicates
a ring surface density
$160\le\sigma_0\le310$ gm/cm$^2$. This finding is consistent with
the the results from the $m=2$ patterns, but this constraint on $\sigma_0$
is again rather loose due to the  $\delta a_{\mbox{\scriptsize edge}}=\pm4$ km 
uncertainty in the ring-edge's semimajor axis.
But our purpose here is to show
how one might use models of free normal modes to infer
the surface density of other rings and narrow ringlets,
which again will likely require knowing the ring-edge's semimajor axis
to $\pm1$ km or better.

Also note that the radial depth of this $m=3$ disturbance is
$\Delta a_{e/10}=50$ km, about three times smaller than the radial depth
of the $m=2$ disturbance.

\pagebreak
\pagebreak

\pagebreak
\pagebreak

\subsection{the free \boldmath{$m=1$} pattern}
\label{m=1}

\begin{figure}[th]
\epsscale{0.86}
\plotone{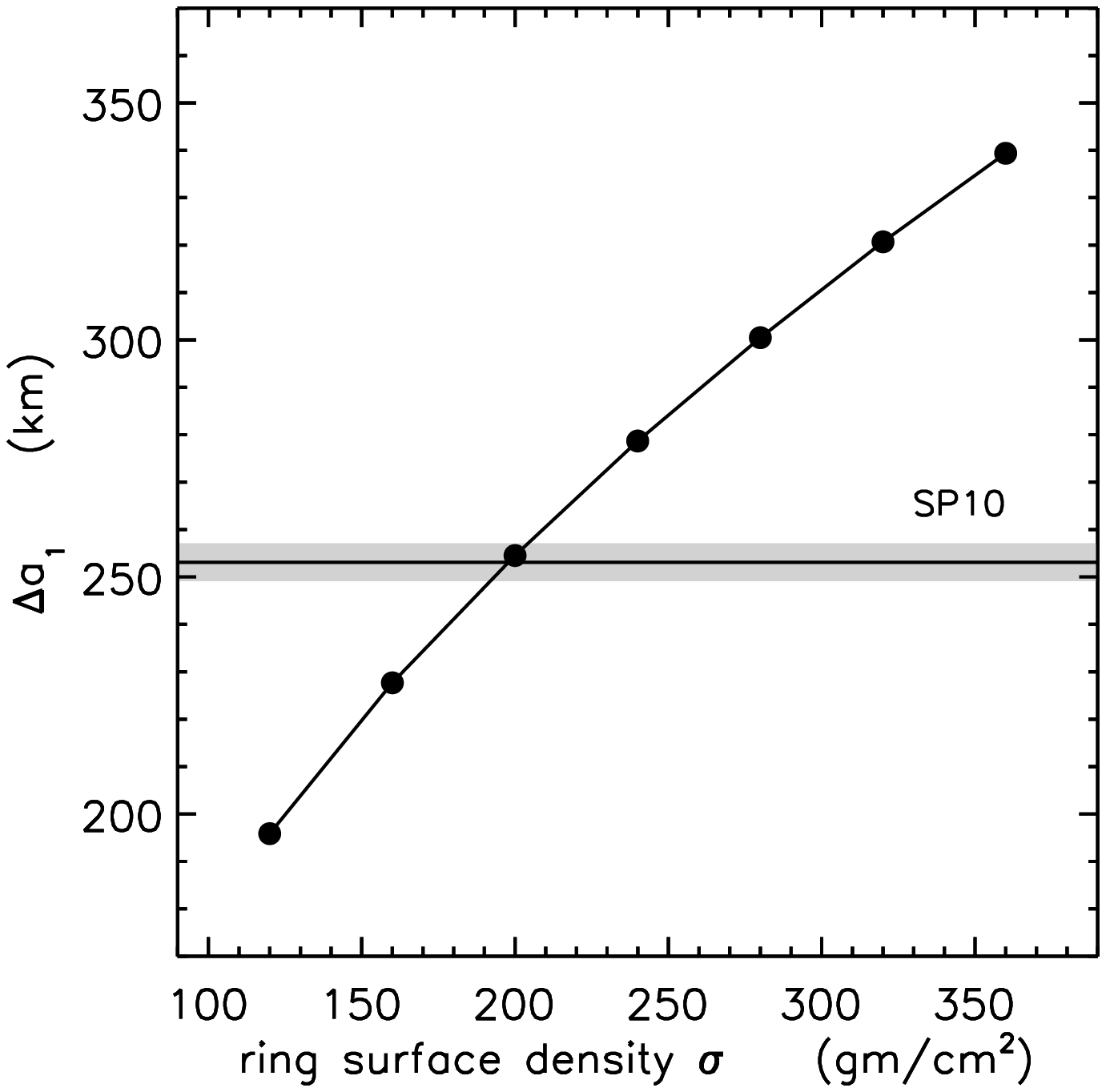}
\figcaption{
    \label{da1_fig}
    Distance $\Delta a_1 = a_{\mbox{\scriptsize edge}} - \tilde{a}_1$
    between the ring's outer edge and the
    inner Lindblad resonance associated with the free normal
    models in the B ring simulations described in Section \ref{m=1},
    plotted versus ring surface density $\sigma_0$. The horizontal line
    is the observed distance with its uncertainty indicated
    by the gray band, from SP10. These simulations use $N_r=100$ streamlines
    with $N_\theta=30$ particles per and $N_rN_\theta=3000$ particles total.
    The streamlines' radial separation is $\Delta a=24.6$ km, and the total
    radial width of the simulated ring is $w=(N_r - 1)\Delta a=2435$ km.\vspace*{-15ex}
}
\end{figure}

The B ring's free $m=1$ mode has an epicyclic amplitude of $\tilde{R}_1=20.9\pm0.4$ km
and a pattern speed $\dot{\tilde{\omega}}_1=5.098\pm0.003$ deg/day that
is slightly faster than the local precession rate, and the
inner Lindblad resonance that is associated with this pattern speed lies 
$\Delta a_1=253\pm4$ km interior to the B ring's outer edge (SP10).
Several simulations of the B ring's $m=1$ pattern are evolved
for model rings having surface densities of $120\le\sigma_0\le360$ gm/cm$^2$.
To excite the $m=1$ pattern at the simulated ring's edge, 
again arrange a fictitious satellite's
orbit so that its $m=1$ ILR lies $\Delta a_1=253$ km
interior to the B ring's edge, which is the site where 
the resonance condition (Eqn.\ \ref{resonance_eqn}) is satisfied
when the satellite's mean angular velocity matches the ring particles'
precession rate, $\Omega_s = \dot{\tilde{\omega}}=\Omega_{ps}$.
The simulated ring is perturbed by a
satellite whose mass is about 20\% that of Mimas,
for $1.6\times10^4$ orbits or 21 years, which excites a forced $m=1$
pattern at the ring's edge that corotates with the satellite. The satellite
is then removed, which converts the forced $m=1$ pattern into a free pattern,
and the ring is evolved for another $6.4\times10^4$ orbits or 83 years. 
For each simulation the free pattern speed is measured, and 
Eqn.\ (\ref{resonance_eqn}) is then used to
convert the free pattern speed into a resonance radius $\tilde{a}_1$, which is displayed
in Fig.\ \ref{da1_fig} that shows that resonance's distance from the ring's outer edge,
$\Delta a_1=a_{\mbox{\scriptsize edge}} - \tilde{a}_1$. As the figure shows,
the free $m=1$ pattern rotates slightly faster in the heavier ring
and thus the associated $m=1$ ILR must lie further inwards in order
to satisfy the resonance condition
$\Omega_{ps}=\dot{\tilde{\omega}}=\frac{3}{2}J_2(R_p/a)^2\Omega$. Again
there is no damping of the free $m=1$ pattern, which stays localized
at the ring's outer edge over the simulation's 83 yr timespan, despite
the simulated ring's viscosity $\nu=100$ cm$^2$/sec.

The radial depth of this $m=1$ disturbance is much greater
than the others, $\Delta a_{e/10}=614$ km, which is about four times larger than 
the $m=2$ disturbance.  Comparing Fig.\ \ref{da1_fig} to
Figs.\ \ref{da2_fig} and \ref{da3_fig} also shows that the LR associated with
the $m=1$ disturbance lies about 10 times further from the ring-edge than the $m=1$ and $m=2$
resonances. Which is why the $m=1$ simulation uses
streamlines whose width $\Delta a$ is $\sim10\times$ larger, since a wider
portion of the B ring-edge must be simulated in order to capture this disturbances'
deeper reach into the B ring.

Note also that the $\pm4$ km uncertainty in this resonance's position relative to
the B ring edge, which is entirely due to the uncertainty in the B ring-edge's
semimajor axis, is in this case relatively small. Which is why the ring's
free $m=1$ mode can also be used to probe its surface density with some precision
(unlike the free $m=2$ and $m=3$ modes),
and is consistent with a B ring surface density of $\sigma_0\simeq200$ gm/cm$^2$,

\pagebreak
\pagebreak

\subsection{convergence tests}
\label{convergence}

A number of simulations have also been performed, which repeat 
the ring simulations using various particle numbers $N_r$ and $N_\theta$
and various widths $w$ of the simulated ring. We find that the results reported here
do not change significantly when the simulated ring is populated 
densely with enough particles,
and when the radial width of the simulated B ring is sufficiently wide.
Those convergence tests reveal that the number of particles
along each streamline must satisfy $N_\theta\ge 20m$, that the radial width of
each streamline should satisfy $\Delta a\le 0.04\Delta a_{e/10}$,
and that the total width of the simulated ring should satisfy $w>4\Delta a_{e/10}$.
All of the simulations reported here satisfy these requirements.

\section{Discussion}
\label{discussion}

This section re-examines the model's treatment of viscous effects at the
ring's edge, and also describes related topics that will be considered
in followup work.

\subsection{the ring's internal forces}
\label{force}

\begin{figure}[th]
\epsscale{0.86}
\plotone{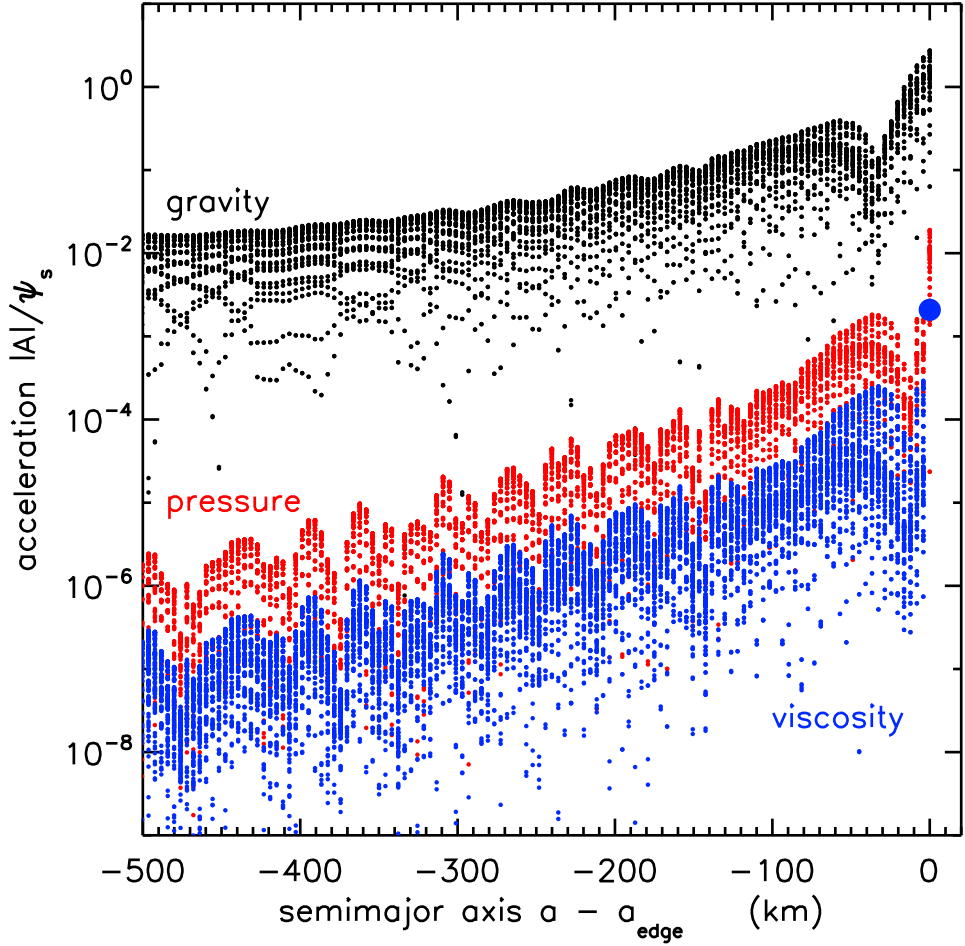}
\figcaption{
    \label{forces_fig}
    The magnitude of the acceleration $|A|$ due to ring self-gravity (black dots),
    pressure (red), and viscosity (blue), is plotted for every particle
    in the B ring simulation having $\sigma_0=200$ gm/cm$^2$  that is described
    in Figs. \ref{R2_fig}--\ref{R_epi_fig}. The ring's internal forces
    are excited by the satellite's periodic forcing, which is conveniently measured by
    the satellite's forcing function $\Psi_s$ (see Eqn.\ 19 of \citealt{HSP09}), and
    these accelerations are displayed in units of $\Psi_s$. 
    This Figure also subtracts from the radial component of $A$ 
    its azimuthally averaged value since that quantity merely changes the
    orbital frequencies $\Omega, \kappa$ slightly without altering the ring's dynamics. 
    Shown are these accelerations
    at time $t=30.9$ yrs when the simulated ring has settled into quasi 
    equilibrium (see Fig.\ \ref{R_epi_fig}). The accelerations are plotted
    versus each streamline's semimajor axis distance from the ring's outer edge,
    $a - a_{\mbox{\scriptsize edge}}$, and these accelerations are periodic in 
    longitude $\theta$, which is why they span a range of values within
    each streamline.
    This simulation also has the viscous acceleration zeroed at the ring's
    outer edge, and the large blue dot on the right indicates the viscous acceleration
    that those particles at the ring's edge would otherwise have experienced; see Section
    \ref{force} for details.\vspace*{-35ex}
}
\end{figure}

Figure \ref{forces_fig} plots the accelerations that the ring's internal
forces---gravity, pressure, and viscosity---exert on each ring particle.
These accelerations are from the nominal $\sigma_0=200$ gm/cm$^2$ simulation 
that is described
in Figs.\ \ref{R2_fig}--\ref{R_epi_fig}, and these accelerations are plotted versus
each particle's distance from the ring's edge,
so those forces obviously get larger closer to the ring's disturbed outer edge.
But the main point of Fig.\ \ref{forces_fig} is that the ring's self gravity
is the dominant internal force in the ring, exceeding the pressure 
force by a factor of $\sim100$ at the
ring's outer edge and by a larger factor elsewhere.
Those pressure forces are also about $\sim10\times$
larger than the ring's viscous forces. But recall that those simulations
had zeroed the viscous acceleration that the ring exerts on
its outermost streamline (Section \ref{B ring}), when that acceleration
should instead be $A_{\nu,\theta}=F/\lambda r$ as indicated by the large blue dot
at the right edge of Fig.\ \ref{forces_fig}. Note though that the neglected
viscous acceleration of the ring's edge is still about $\sim1000\times$
smaller than that due to ring gravity and $\sim10\times$ smaller than that due to
ring pressure. So this justifies neglecting, at least for the short-term $t\sim100$ yr
simulations considered here, the much smaller viscous forces at the ring's outer edge.

Nonetheless this study's neglect of the small viscous force at the ring's outer
edge implies that this model does not yet account for the B ring's 
radial confinement by Mimas' $m=2$ ILR. So there appears to be some missing physics
that will be necessary if one is interested in the ring's resonant confinement
or the ring's long-term evolution over $t\gg100$ yr timescales. The suspected
missing physics is described below.

\pagebreak
\pagebreak

\subsubsection{unmodeled effects: the viscous heating of a resonantly confined ring-edge}
\label{viscous heating}

The model's inability to confine the B ring's outer
edge at Mimas' $m=2$ ILR may be a consequence
of the ring's kinematic viscosity $\nu$ being treated here as
a constant parameter everywhere in the simulated ring. Although treating $\nu$ as a constant
is a simple and plausible way to model the effects of the ring's viscous friction, it
might not be adequate or accurate if one wishes to simulate the resonant confinement
of a planetary ring. This is because the ring's viscosity transports 
both energy and angular momentum radially outwards through the ring. So if the ring's outer
edge is to be confined by a satellite's $m^{\mbox{\scriptsize th}}$ Lindblad resonance,
the satellite must absorb the ring's outward angular momentum flux,
which it can do by exerting a negative gravitational torque at the ring's edge. But 
\cite{BGT82} show via a simple Jacobi-integral argument 
that resonant interactions only allow the satellite
to absorb but a fraction of the energy that viscosity delivers to the ring-edge.
Consequently the ring's viscous friction still delivers some
orbital energy to the ring-edge where
it accumulates and heats up the ring particles' random velocities $c$.
And if collisions among particles are the main source of the ring's viscosity,
then $\nu_s\simeq c^2\tau/2\Omega(1+\tau^2)$ where $\tau\propto\sigma$ is the
ring's optical depth \citep{GT82}. In this case viscous heating
would increase $c$ as well as $\nu_s$ at the ring's edge. The enhanced dissipation there
should also increase the angular lag $\phi$ between the
ring-edge's forced pattern and the satellite's
longitude (see Eqn.\ 83b of \citealt{HSP09}). Which will also be important because
the gravitational torque that the satellite exerts on the ring-edge varies as
$\sin\phi$ \citep{HSP09}, and that torque needs to be boosted if the satellite
is to confine the spreading ring.

To model this phenomenon properly, the {\tt epi\_int} code
also needs to employ an energy equation,
one that accounts for how viscous heating tends to increase the ring particles'
dispersion velocity $c$ and viscosity $\nu_s$ nearer the ring's edge. The increased
dissipation and the resulting orbital lag will allow the satellite
to exert a greater torque on the ring which, we suspect, will enable the satellite
to resonantly confine the simulated ring's outer edge. The derivation of this energy equation
and its implementation in {\tt epi\_int} are ongoing, and those results will be reported
on in a followup study.

\subsection{an alternate equation of state}
\label{EOS}

The EOS adopted here is appropriate 
for a dilute gas of colliding ring particles whose
mutual separations greatly exceed their sizes. 
This should be regarded as a limiting case since ring particles can of course
be packed close to each other in the ring. But \cite{BGT85}
consider the other extreme limiting case, with
close-packed particles that reside shoulder to shoulder in the ring. In that
case the ring is expected to behave as an incompressible fluid
whose volume density $\rho=\sigma/2h$ stays constant.
So when some perturbation causes ring streamlines to bunch up
and increases the ring's surface density $\sigma$, the ring's vertical
scale height $h$ also increases as ring particles are forced
to accumulate along the vertical
direction. This in turn increases the ring's pressure due to the
larger gravitational force along the vertical direction.

\cite{BGT85} show that infinitesimal density waves in an
incompressible disk are unstable and grow in amplitude over time.
This phenomenon is related to the viscous overstability,
and \cite{LR95} show that it can
distort a narrow eccentric ringlet's streamlines in a way that
accounts for its $m=1$ shapes. \cite{BGT85} also suggest
that unstable density waves can be trapped between a Lindblad resonance
and the B ring's outer edge, which might explain the normal modes
seen there, and \cite{SP10} use this concept to estimate the ring's
surface density there.

But keep in mind that this instability only occurs when the ring particles are
densely packed to the point of being incompressible, which requires the ring to
be very thin and dynamically cold. We have shown here that 
the amplitude of the B ring's forced motions indicates
that the ring-edge has a surface density
$\sigma\simeq200$ gm/cm$^2$.  So if this ring is incompressible and composed
of icy spheres having a mean volume density of $\rho=\sigma/2h\simeq0.5$ gm/cm$^3$,
this then requires a B ring thickness of only $h\sim2$ meters, which is rather thin compared
to other estimates \citep{CBC10}. Similarly  the ring particles' dispersion velocity $c$
must be small compared to that expected for a dilute particle gas,
so $c\ll (h\Omega\sim0.3$ mm/sec), which again is cold compared to
all other estimates for Saturn's rings \citep{CBC10}. The upshot is that
an incompressible EOS requires the ring to be very thin and dynamically cold,
likely much colder and thinner than is generally thought. Consequently
we are optimistic that the compressible EOS used here, $p=\sigma c^2$,
is the appropriate choice for simulations
of the outer edge of Saturn's B ring. Nonetheless
in a followup investigation we do intend
to encode the incompressible EOS into {\tt epi\_int},
to see if the BGT instability can account for the higher $m\ge2$ free modes
that are seen at the outer edge of the B ring and in many other narrow ringlets.

\subsection{impulse origin for normal modes}
\label{impulse}

The simulations of Section \ref{B ring} used a fictitious temporary satellite 
to excite the free modes that occur at many Saturnian
ring edges. These simulations used an admittedly ad hoc method---the sudden
appearance and disappearance of a satellite---to excite these modes.
Nonetheless these models demonstrate that transient and impulsive events
can excite normal modes at ring edges, and those simulations show that
normal modes can persist at the ring's edge for hundreds of years after
the disturbance has occurred. Which suggests that an
impulsive event in the recent past, 
perhaps an impact into Saturn's rings, might be responsible
for exciting the normal modes that are seen at the outer edge of the B ring,
as well as the normal modes that are also seen along the edges of
several narrower ringlets \citep{FMR10, HNB10, FNC11, NFH12}

The possibility that normal modes are due to an impact
is  motivated by the discovery of vertical corrugations
in Saturn's C and D rings \citep{HBS07, HBE11} and in Jupiter's main dust ring 
\citep{SHB11}. These vertical structures are spirals that span
a large swath of each ring, and they are observed to wind up over time due to
the central planet's oblateness. Evolving the vertical corrugations 
backwards in time also unwinds their spiral pattern until some moment when the
affected region is a single tilted plane. Unwinding the Jovian corrugation
shows that that disturbance occurred very close to the date when the
tidally disrupted comet Shoemaker-Levy 9
impacted Jupiter in 1994, which suggests an impact from a tidally disrupted
comet as the origin of these ring-tilts \citep{SHB11}. However
a single sub-km comet fragment cannot tilt a large $\sim2\times10^5$ km-wide
planetary ring. But a disrupted comet can produce an extended cloud
of dust, and if that disrupted dust cloud returns to the planet
with enough mass and momentum, then it might tilt a ring that
at a later date would be observed as a spiral corrugation.

However the tidal disruption of comet
about a low-density planet like Saturn is more problematic, because tidal
disruption only occurs when the comet's orbit is truly close
to parabolic and not too hyperbolic, and
with periapse just above the planet's atmosphere \citep{ST92, RBL98}.

But it is easy to envision an alternate scenario that might be more likely,
with a small km-sized comet originally in a heliocentric orbit coming
close enough to Saturn to instead strike the main A or B rings. This scenario is 
more probable because the cross-section available to orbits impacting the main rings
is significantly larger than those resulting in tidal disruption.
The impacting comet's considerably greater
momentum will nonetheless carry the
impactor through the dense A or B rings, but the collision itself is likely
energetic enough to shatter the comet. And if that collision is
sufficiently dissipative, then the resulting cometary debris
will then stay bound to Saturn, and in an orbit 
that will return that debris back into the ring system on its next orbit.
Small differences among the orbits of individual debris particles'
means that, when the debris encounters the rings again,
that impacting debris will be spread
across a much larger footprint on the ring, which presumably will 
allow any dense rings or ringlets to absorb the debris' mass and momentum in a way that
effectively gives the ring particles there a sudden velocity kick $\mathbf{\Delta v}$
in proportion to the comet debris density $\rho$ and velocity
$\mathbf{v}_r$ relative to the ring matter. But if comet Shoemake-Levy 9's (SL9)
impact with Jupiter is any guide, then impact by a cloud of
comet debris could last as long week of time, which might tend to smear this effect out
due to the ring's orbital motion.
But that effect would be offset if the debris train's dust cloud is also rather clumpy,
like the SL9 debris train was. Indeed, it is possible that this scenario might also
account for the spiral corrugations of Saturn's C and D ring.
It is also conceivable that an inclined cloud of impacting comet
debris might also excite the vertical analog of normal modes---long-lived
vertical oscillations of a ring's edge. This admittedly speculative scenario will be
pursued in a followup study, to determine whether debris from an impact-disrupted
comet can excite the normal modes seen at ring edges, and to
determine the mass of the progenitor comet that would be needed
to account for these modes' observed amplitudes.

\section{Summary of results}
\label{summary}

We have developed a new N-body integrator that
calculates the global evolution of a self-gravitating planetary ring
as it orbits an oblate planet. The code is called
{\tt epi\_int}, and it uses the same
kick-drift-step algorithm as is used in other symplectic integrators
such as {\tt SYMBA} and {\tt MERCURY}. However the velocity kicks that are
due to ring gravity are computed via an alternate method
that assumes that all particles inhabit a discreet number of streamlines
in the ring.  The use of streamlines to calculate ring self gravity 
has been used in analytic studies of rings \citep{GT79, BGT83aj, BGT86},
and the streamline concept is easily implemented in an N-body code. A 
streamline is the closed  path through the ring that is traced by
particles having a common semimajor axis.
All streamlines are radially close to each other, so the gravitation
acceleration due to a streamline is simply that due to a long wire,
$A=2G\lambda/\Delta$ where $\lambda$ is the streamline's linear density and
$\Delta$ is the particle's distance from the streamline. 
Which is very useful since particles are responding to the pull of smooth
wires rather than discreet clumps of ring matter
so there is no gravitational scattering. Which means that only a modest number
of particles are needed, typically a few thousand, to simulate all $360^\circ$
of a scalloped ring like the outer edges of Saturn's A and B ring.
Only a few thousand particles are also needed to simulate linear as well as
nonlinear spiral density waves,
and execution times are just a few hours on a desktop PC. 

Another distinction occurs during the particles' unperturbed drift step
when particles follow the epicyclic orbit of \cite{LR95} about an oblate planet, 
rather than the usual Keplerian orbit about a spherical planet. This effectively
moves the perturbation due to the planet's oblate figure out of the
integrator's kick step and into the drift step. The code
also employs hydrodynamic pressure and viscosity
to account for the transport of linear and angular momentum through the ring 
that arises from collisions among ring particles. Another convenience of the
streamline formulation is that it easily accounts for the large pressure drop
that occurs at a ring's sharp edge, as well as the large viscous torque that the
ring exerts there. The model also accounts for the mutual gravitational
perturbations that the ring and the satellites exert on each other.
The {\tt epi\_int}  code is written in IDL, and the source code is available for download at
\href{http://gemelli.spacescience.org/~hahnjm/software.html}
    {\tt http://gemelli.spacescience.org/\char`\~hahnjm/software.html}.

This integrator is used to simulate the forced response that
the satellite Mimas excites at its $m=2$ inner Lindblad resonance (ILR) 
that lies near the outer edge of Saturn's B ring.
That resonance lies $\Delta a_2=12 \pm4$ km inwards of the ring's
edge, and simulations show that the ring's forced epicyclic amplitude
varies with the ring's surface density $\sigma_0$ as $R_2\propto\sigma_0^{0.67}$.
Good agreement with Cassini measurements of $R_2$ occurs when
the simulated ring has a surface density of $\sigma_0=195\pm 60$ gm/cm$^2$
(see Fig.\ \ref{R2_fig}), where the uncertainty in $\sigma_0$
is dominated by the $\delta a_{\mbox{\scriptsize edge}}=4$ km 
uncertainty that \cite{SP10} find
in the ring-edge's semimajor axis. This $\sigma_0$ is the mean surface density
over that part of the B ring that is disturbed by this resonance, whose 
influence in the ring extends to a radial distance of
$\Delta a_{e/10}\sim150$ km from the B ring's outer edge.
And if we naively assume that this surface density is the same
everywhere  across Saturn's B ring, then its total mass is 
about $90\%$ of Mimas' mass.

Cassini observations reveal that the outer edge of Saturn's B ring
also has several free normal modes that are not excited by any
known satellite resonances. Although the mechanism that excites these free modes
is uncertain, we are nonetheless able to excite free modes in a simulated ring via 
various ad-hoc methods. For instance, a fictitious
satellite's $m^{\mbox{\scriptsize th}}$ Lindblad resonance 
is used to excite  a forced pattern at the ring edge. Removing that satellite then
converts the forced patten into a free normal mode that persists in
these simulations for up to $\sim100$ years or $\sim10^5$ orbits
without any damping, despite the simulated ring having a kinematic viscosity
of $\nu=100$ cm$^2$/sec; see Fig.\ \ref{m3_fig} for one example.

Alternatively, starting the ring particles in circular
orbits while subject to Mimas' $m=2$ gravitational perturbation excites both a forced
and a free $m=2$ pattern that initially null each other precisely at the start of the
simulation. But the forced patten corotates with Mimas' longitude while the free
pattern rotates slightly faster in a heavier ring, which suggests that
a free mode's pattern speed
can also be used to infer a ring's surface density $\sigma_0$. However
the free pattern speed is also influenced by the $J_4$ and higher terms
in the oblate planet's gravity field, which are absent from this model which
only accounts for the $J_2$ component.
So the simulated pattern speed cannot be compared directly to the observed
pattern speed; see Fig.\ \ref{m2_fig}. To avoid this difficulty,
the resonance condition (Eqn.\ \ref{resonance_eqn}) is used
to calculate the radius of the Lindblad resonance
that is associated with the free normal mode. Plotting the distances
of the simulated and observed resonances from the B ring's edge
(Figs.\ \ref{da2_fig}, \ref{da3_fig}, and \ref{da1_fig})
then provides a convenient way to compare simulations to observations
of free modes in a way that is insensitive to the planet's oblateness.

Simulations of the B ring's free $m=2$ and $m=3$ patterns are consistent with
Cassini measurements of the B ring's normal modes when
the simulated ring-edge again has a surface density of $\sigma_0\sim200$ gm/cm$^2$,
which is a nice consistency check. But these particular measurements
do not provide tight constraint on the ring's $\sigma_0$,
due to the fact that the $m=2$ and $m=3$ Lindblad resonances only lie
$\Delta a_m\sim25$ km from  the outer edge of a ring whose semimajor axis $a$
is uncertain by $\delta a_{\mbox{\scriptsize edge}}=4$ km.
However the B ring's free $m=1$
normal mode does lie much deeper in the ring's interior, $\Delta a_1=253\pm4$, 
so the uncertainly in its location is fractionally much smaller, and this 
normal mode does confirm the $\sigma_0\simeq200$ gm/cm$^2$
value that was inferred from simulations of the B ring's forced response $R_2$.

One of the goals of this study is to determine whether simulations of free modes
can be used to determine the surface density and mass of a narrow ringlet.
Such ringlets show a broad spectrum of free normal models over $0\le m\le 5$
\citep{FMR10, HNB10, FNC11, NFH12}, and the answer appears to be yes since
free pattern speeds do increase with $\sigma_0$.
However Section \ref{free m=2} shows
that the semimajor axes of the ringlet's edges likely
need to be known to a precision of 
$\delta a_{\mbox{\scriptsize edge}}\sim1$ km in order for a
free mode to provide a useful measurement of the ringlet's $\sigma_0$.

The origin of these free modes, which are quite common along the edges
of Saturn's broad rings and its many narrow ringlets, is uncertain.
\cite{BGT85} show that, if a planetary ring's particles are packed 
shoulder to shoulder such that the ring behaves like an incompressible fluid,
then that ring is unstable to the growth of density waves, a phenomenon also termed
viscous overstability, and they suggest that
the B ring's normal modes might be due to unstable waves 
that are trapped between a Lindblad
resonance and the ring's edge. To study this further, we will in a followup study adapt
{\tt epi\_int} to employ an incompressible equation of state,
to see if the viscous overstability can in fact account for the free normal
modes seen along the Saturnian ring edges.

Although the current version of {\tt epi\_int} does not account for the
origin of these free modes, one can still
plant a free mode along the edge of a simulated
ring by temporarily perturbing a ring at a fictitious satellite's Lindblad resonance,
and then removing that satellite, which creates an unforced mode
that persists undamped at the ring-edge 
for more than $\sim10^5$ orbits or $\sim100$ yrs despite the simulated ring having a
kinematic viscosity of $\nu=100$ cm$^2$/sec. Because this forcing is suddenly
turned on and off, this suggests that any sudden
or impulsive disturbance of the ring can excite normal modes, with those
disturbances possibly
persisting for hundreds or maybe thousands of years.
And in Section \ref{impulse} we suggest that the Saturnian normal modes
might be excited by an impact with a collisionally disrupted cloud of
comet dust. This is a slight variation of the scenario that 
\cite{HBS07} and \cite{SHB11} propose for the origin of corrugated planetary
rings, and in a followup investigation we intend to determine whether such
impacts can also account for the normal modes seen in Saturn's rings.

And lastly, we find that {\tt epi\_int}'s treatment of ring viscosity
has difficulty accounting for the
radial confinement of the B ring's outer edge by Mimas' $m=2$ inner Lindblad
resonance. This model employs a kinematic shear viscosity $\nu_s$ that is
everywhere a constant, which causes the simulation's outermost streamline
to slowly but steadily drift radially outwards. Which in turn causes the ring's forced
epicyclic amplitude $R_2$ to slowly grow over time, and makes difficult any comparison
to Cassini's measurement of $R_2$. To sidestep this difficulty,
the model zeros the torque that the simulated ring exerts on its outermost streamline,
which does allow the ring to settle into a static configuration that
can be compared to Cassini observations and yields a measurement of the ring's 
surface density $\sigma_0$.
This approximate treatment is also examined in in Section \ref{force}, which
shows that the viscous acceleration of the ring-edge, had it
been included in the simulation, 
is still orders of magnitude smaller than that due to ring self gravity.
So this study of the dynamics of the B ring's forced 
and free modes is not adversely impacted by this approximate treatment.
But this does mean that the B ring's radial confinement is still an unsolved
problem, and Section \ref{viscous heating} suggests that this might be
a consequence of treating $\nu_s$ as a constant. \cite{BGT82}
show that viscosity's outward
transport of energy should also heat the ring's outer edge and
increase the ring particles' dispersion velocity $c$ there.
And if collisions among ring particles are the dominant source of ring viscosity,  
then $\nu_s\propto c^2$ and viscous dissipation would be enhanced
at the ring edge, which in turn would increase the angular lag between
the ring's forced response and the Mimas' longitude. That then would
increase the gravitational torque that that satellite exerts on the ring-edge.
So in a followup study we will modify {\tt epi\_int} to address this problem
in a fully self-consistent way, to see if enhanced
dissipation at the ring-edge also increases Mimas' gravitational torque there
sufficiently to prevent the B ring's outer edge from
flowing viscously beyond that satellite's
$m=2$ inner Lindblad resonance.

\acknowledgments

\begin{center}
  {\bf Acknowledgments}
\end{center}

J.\ Hahn's contribution to this work was supported by grant NNX09AU24G
issued by NASA's Science Mission Directorate via its Outer Planets
Research Program. The authors thank Denise Edgington of the University of
Texas' Center for Space Research (CSR) for composing Fig.\ \ref{streamline_fig},
and J.\ Hahn thanks Byron Tapley for graciously providing office space 
and the use of the facilities at CSR. The authors are also grateful for the helpful
suggestions provided by an anonymous reviewer.

\appendix
\section{Appendix \ref{shear_appendix}}
\label{shear_appendix}

The following calculates the flux of angular momentum that is
communicated via a disk's viscosity. The disk is flat and thin
and has a vertical halfwidth $h$ and constant volume density $\rho$ that is
related to its surface density $\sigma$ via $\rho=\sigma/2h$. The disk is assumed
viscous, and its gravity is ignored here since this Appendix is only interested
in the angular momentum flux that is transported solely by viscosity.

The density of angular momentum in the disk 
is $\bm{\ell} = \mathbf{r}\times\rho\mathbf{v}$, and the
vertical component along the $z=x_3$ axis is $\ell_3 = x_1\rho v_2 - x_2\rho v_1$
in Cartesian coordinates $x=x_1$ and $y=x_2$ where $\rho$ and $v_i$ are functions
of position and time, so the time rate of change of $\ell_3$ is
\begin{eqnarray}
    \label{dl3/dt}
    \frac{\partial\ell_3}{\partial t} &= x_1\displaystyle\frac{\partial}{\partial t}(\rho v_2)
        - x_2\displaystyle\frac{\partial}{\partial t}(\rho v_1).
\end{eqnarray}
The time derivatives in the above are Euler's equation,
\begin{eqnarray}
    \label{EEqn}
    \frac{\partial}{\partial t}(\rho v_i) &=-\displaystyle\sum_{k=1}^3
        \frac{\partial\Pi_{ik}}{\partial x_k}
\end{eqnarray}
where the $\Pi_{ik}$ are the elements of the momentum flux density tensor
\begin{eqnarray}
    \label{Pi}
    \Pi_{ik} &= \rho v_iv_k + \delta_{ik}p - \sigma'_{ik}
\end{eqnarray}
where $p$ is the pressure and $\sigma'_{ik}$ are the elements of the viscous stress tensor
\citep{LL87}. Inserting Eqn.\ (\ref{Pi}) into (\ref{dl3/dt}) yields
\begin{eqnarray}
    \label{dl3/dt_v2}
    \frac{\partial\ell_3}{\partial t} &= -x_1\nabla\cdot\bm{\Pi}_2 + x_2\nabla\cdot\bm{\Pi}_1
\end{eqnarray}
where the vector
\begin{eqnarray}
    \label{Pi_vector}
    \bm{\Pi}_i=\displaystyle\sum_{k=1}^3\Pi_{ik}\bm{\hat{x}}_k
\end{eqnarray}
is the flux density of the $i$ component of
linear momentum and $\bm{\hat{x}}_k$ is the unit vector along
the $x_k$ axis. Equation (\ref{dl3/dt_v2}) can be rewritten
\begin{eqnarray}
    \label{dl3/dt_v3}
    \frac{\partial\ell_3}{\partial t} &= -\nabla\cdot(x_1\bm{\Pi}_2 - x_2\bm{\Pi}_1)
        +\bm{\Pi}_2\cdot\nabla x_1 - \bm{\Pi}_1\cdot\nabla x_2
\end{eqnarray}
but note that $\bm{\Pi}_1\cdot\nabla x_2 - \bm{\Pi}_2\cdot\nabla x_1=\Pi_{21}-\Pi_{12}=
\sigma'_{12} - \sigma'_{21}=0$ since the viscous stress tensor is symmetric
(Eqn.\ \ref{sigma'}), so
\begin{eqnarray}
   \label{dl3/dt_v4}
   \frac{\partial\ell_3}{\partial t} &= -\nabla\cdot \bm{F}_3
\end{eqnarray}
where
\begin{eqnarray}
   \label{F3}
   \bm{F}_3 &= x_1\bm{\Pi}_2 - x_2\bm{\Pi}_1.
\end{eqnarray}
Integrating Eqn.\ (\ref{dl3/dt_v4}) over some volume $V$ that is bounded by area $A$ yields
\begin{eqnarray}
   \label{F3_flux}
   \frac{\partial}{\partial t}\int_V \ell_3 dV &= -\int_V\nabla\cdot \bm{F}_3 dV =
        -\int_A \bm{F}_3\cdot \bm{dA}
\end{eqnarray}
by the divergence theorem, so Eqn.\ (\ref{F3_flux}) indicates that
$\bm{F}_3$ is the flux of the $x_3$ component
of angular momentum out of volume $V$ that is being
transported by advection, pressure, and viscous effects.

This Appendix is interested in the part of $\bm{F}_3$ that is due to
viscous effects, which will be identified as $\bm{F}'_3$ and
is obtained by replacing $\Pi_{ik}$ in Eqn.\ (\ref{Pi}) with $-\sigma'_{ik}$ so
\begin{eqnarray}
   \label{F3'}
   \bm{F}'_3 &=  (x_2\sigma'_{11} - x_1\sigma'_{21})\bm{\hat{x}}_1
                + (x_2\sigma'_{12} - x_1\sigma'_{22})\bm{\hat{x}}_2.
\end{eqnarray}
This is the 2D flux of the $x_3$ component of angular momentum 
that is transported by the disk's viscosity whose horizontal components in
Cartesian coordinates are $\bm{F}'_3=F'_1\bm{\hat{x}}_1 + F'_2\bm{\hat{x}}_2$
where $F'_1=x_2\sigma'_{11} - x_1\sigma'_{21}$ and
$F'_2=x_2\sigma'_{12} - x_1\sigma'_{22}$. However this Appendix desires
the radial component of $\bm{F}'_3$ are some site $r,\theta$ in the disk,
which is $F'_r=F'_1\cos\theta + F'_2\sin\theta$.

The elements of the viscous stress tensor are \citep{LL87}
\begin{eqnarray}
    \label{sigma'}
    \sigma'_{ik} &= \eta\left(\displaystyle\frac{\partial v_i}{\partial x_k} +
        \frac{\partial v_k}{\partial x_i}\right)
        + (\zeta - \frac{2}{3}\eta)\delta_{ik}\nabla\cdot\bm{v}
\end{eqnarray}
where $\eta$ is the shear viscosity, $\zeta$ is the bulk viscosity,  
and $\delta_{ik}$ is the Kronecker delta. Inserting this into $F'_r$
and replacing $x_1=r\cos\theta$ and $x_2=r\sin\theta$
then yields
\begin{eqnarray}
   \label{F'_r}
   F'_r &=-\eta r\left(\displaystyle\frac{\partial v_1}{\partial x_2}
        + \frac{\partial v_2}{\partial x_1}\right)\cos2\theta +
         \eta r\left(\displaystyle\frac{\partial v_1}{\partial x_1}
        - \frac{\partial v_2}{\partial x_2}\right)\sin2\theta.
\end{eqnarray}
The horizontal velocities are $v_1=v_r\cos\theta  - v_\theta\sin\theta $
and $v_2 = v_r\sin\theta  + v_\theta\cos\theta $ when written in terms 
of their radial component $v_r$ and tangential component $v_\theta=r\dot{\theta}$.
The derivatives in Eqn.\ (\ref{F'_r}) are
\begin{eqnarray}
    \label{dv/dx}
    \begin{split}
            \frac{\partial v_1}{\partial x_1} &=
                \left(\cos\theta\frac{\partial}{\partial r}
                      -\frac{\sin\theta}{r}\frac{\partial}{\partial \theta}\right)v_1\\
            &= \cos^2\theta\frac{\partial v_r}{\partial r} 
                - \sin\theta\cos\theta r\frac{\partial\dot{\theta}}{\partial r}
                + \frac{\sin^2\theta}{r}v_r
                - \frac{\sin\theta\cos\theta}{r}\frac{\partial v_r}{\partial\theta}
                + \frac{\sin^2\theta}{r}\frac{\partial v_\theta}{\partial\theta}\\
            \frac{\partial v_2}{\partial x_2} &=
                \left(\sin\theta\frac{\partial}{\partial r}
                      +\frac{\cos\theta}{r}\frac{\partial}{\partial \theta}\right)v_2\\
            &= \sin^2\theta\frac{\partial v_r}{\partial r} 
                + \sin\theta\cos\theta r\frac{\partial\dot{\theta}}{\partial r}
                + \frac{\cos^2\theta}{r}v_r
                + \frac{\sin\theta\cos\theta}{r}\frac{\partial v_r}{\partial\theta}
                + \frac{\cos^2\theta}{r}\frac{\partial v_\theta}{\partial\theta}\\
            \frac{\partial v_1}{\partial x_2} &=
                \left(\sin\theta\frac{\partial}{\partial r}
                  +\frac{\cos\theta}{r}\frac{\partial}{\partial \theta}\right)v_1\\
            \frac{\partial v_2}{\partial x_1} &=
                \left(\cos\theta\frac{\partial}{\partial r}
                  -\frac{\sin\theta}{r}\frac{\partial}{\partial \theta}\right)v_2
    \end{split}
\end{eqnarray}
when written in terms of cylindrical coordinates, and the combinations of derivatives
in Eqn.\ (\ref{F'_r}) are 
\begin{mathletters}
    \label{dv/dx +- dv/dx}
    \begin{eqnarray}
        \label{dv1/dx2 + dv2/dx1}
        \frac{\partial v_1}{\partial x_2} + \frac{\partial v_2}{\partial x_1}
            &=& \left(\frac{\partial v_r}{\partial r} 
                -\frac{1}{r}\frac{\partial v_\theta}{\partial \theta}
                - \frac{v_r}{r}\right)\sin2\theta
                + \left(\frac{\partial v_\theta}{\partial r}
                +\frac{1}{r}\frac{\partial v_r}{\partial \theta}
                - \frac{v_\theta}{r}\right)\cos2\theta\\
        \label{dv1/dx1 - dv2/dx2}
        \frac{\partial v_1}{\partial x_1} - \frac{\partial v_2}{\partial x_2}
            &=& \left(\frac{\partial v_r}{\partial r} 
                -\frac{1}{r}\frac{\partial v_\theta}{\partial \theta}
                - \frac{v_r}{r}\right)\cos2\theta
                - \left(\frac{\partial v_\theta}{\partial r}
                +\frac{1}{r}\frac{\partial v_r}{\partial \theta}
                - \frac{v_\theta}{r}\right)\sin2\theta,
    \end{eqnarray}
\end{mathletters} 
Inserting these into Eqn.\ (\ref{F'_r}) then yields a result that is
thankfully much more compact,
\begin{eqnarray}
   \label{F'_r2}
   F'_r &=-\eta\left( \displaystyle r^2\frac{\partial\dot{\theta}}{\partial r} + 
        \frac{\partial v_r}{\partial \theta}\right)
        \simeq -\eta r^2\displaystyle \frac{\partial\dot{\theta}}{\partial r},
\end{eqnarray}
noting that the second term in Eqn.\ (\ref{F'_r2}) may be neglected since the azimuthal
gradient is much smaller than the radial gradient for the disks considered here.
This is the radial component of the disk's 2D viscous angular momentum flux density,
so the 1D viscous angular momentum flux density is Eqn.\ (\ref{F'_r2}) integrated through
the disk's vertical cross section:
\begin{eqnarray}
    \label{F_app}
    F &= \int_{-h}^{h} F'_rdx_3=
       -\nu_s\sigma r^2\displaystyle \frac{\partial\dot{\theta}}{\partial r}
\end{eqnarray}
where $\nu_s=\eta/\rho$ is the disk's kinematic shear viscosity.

\section{Appendix \ref{bulk_appendix}}
\label{bulk_appendix}

The flux density of $x_1$-type momentum is $\bm{\Pi}_1$ (see Eqn.\ \ref{Pi_vector}) while the 
flux density of $x_2$-type momentum is $\bm{\Pi}_2$, so the flux density of
radial momentum is $\bm{G} = \cos\theta\bm{\Pi}_1 + \sin\theta\bm{\Pi}_2$ and the
radial component of this momentum flux density is
\begin{eqnarray}
    \label{G_appendix}
    G_r &= \bm{G\cdot\hat{r}} = (\cos\theta\Pi_{11} + \sin\theta\Pi_{21})
        \bm{\hat{x_1}\cdot\hat{r}}
        + (\cos\theta\Pi_{12} + \sin\theta\Pi_{22})\bm{\hat{x_2}\cdot\hat{r}}\\
    &= \cos^2\theta\Pi_{11} + \sin\theta\cos\theta(\Pi_{12} + \Pi_{21}) + \sin^2\theta\Pi_{22}
\end{eqnarray}
where $\bm{\hat{r}}$ is the unit vector in the radial direction. The part of that
momentum flux that is transported solely by viscous effects will be called $G_r'$
and is again obtained by replacing the $\Pi_{ik}$ in the above
with $-\sigma'_{ik}$:
\begin{eqnarray}
    \label{G'_r}
    G_r'&=& -\cos^2\theta\sigma_{11} - \sin\theta\cos\theta(\sigma'_{12} + \sigma'_{21}) 
        - \sin^2\theta\sigma'_{22}\\
        &=& - 2\eta\left[   \displaystyle\cos^2\theta\frac{\partial v_1}{\partial x_1}
                         + \sin^2\theta\frac{\partial v_2}{\partial x_2} 
                         + \sin\theta\cos\theta \left( 
                           \frac{\partial v_1}{\partial x_2} 
                         + \frac{\partial v_2}{\partial x_1}\right)\right]
            - (\zeta - \frac{2}{3}\eta)\bm{\nabla\cdot v}.
\end{eqnarray}

Equations (\ref{dv/dx}) provide the combination 
\begin{eqnarray}
        \begin{split}
              \cos^2\theta\frac{\partial v_1}{\partial x_1} 
            + \sin^2\theta\frac{\partial v_2}{\partial x_2} &=
            \left(\frac{3}{4} + \frac{1}{4}\cos4\theta\right)\frac{\partial v_r}{\partial r}
            -\frac{1}{4}\sin4\theta r\frac{\partial \dot{\theta}}{\partial r}
            +\frac{1}{2r}\sin^22\theta v_r
            -\frac{1}{4r}\sin4\theta \frac{\partial v_r}{\partial \theta}\\
            &+\frac{1}{2r}\sin^22\theta \frac{\partial v_\theta}{\partial \theta},
        \end{split}
\end{eqnarray}
and inserting this plus Eqn.\ (\ref{dv1/dx2 + dv2/dx1}) into Eqn.\ (\ref{G'_r}) then yields
\begin{eqnarray}
    \label{G'_r2}
    G_r' &= -\displaystyle\left(\frac{4}{3}\eta +\zeta\right)\frac{\partial v_r}{\partial r}
            -\left(\zeta-\frac{2}{3}\eta\right)
             \left(\frac{v_r}{r} + \frac{1}{r}\frac{\partial v_\theta}{\partial \theta}\right)
\end{eqnarray}
but the $\partial v_\theta/\partial \theta$ term 
is again neglected in the streamline approximation.
This is the 2D radial momentum flux due to viscous transport, so
the vertically integrated linear momentum flux due to viscosity is
\begin{eqnarray}
    \label{G_appendix2}
    G &=& \int_{-h}^{h} G'_rdx_3=
        -\displaystyle\left(\frac{4}{3}\nu_s +\nu_b\right)\sigma\frac{\partial v_r}{\partial r}
            -\left(\nu_b-\frac{2}{3}\nu_s\right)\frac{\sigma v_r}{r}.
\end{eqnarray}

\bibliography{biblio}

\end{document}